\documentclass[10pt,letterpaper, twocolumn]{revtex4}
\usepackage[T1]{fontenc}
\usepackage{amsmath}
\usepackage{amsthm}
\usepackage{amsfonts}
\usepackage{amssymb}
\usepackage{graphicx}
\usepackage{dsfont}
\usepackage{hyperref}
\usepackage{bbm}

\newcommand{\ha}{\frac{1}{2}}

\newcommand{\N}{\mathbb{N}}

\newcommand{\R}{\mathbb{R}}

\newcommand{\ev}[1]{\langle#1\rangle}
\newcommand{\bra}[1]{\langle#1|}
\newcommand{\ket}[1]{|#1\rangle}

\newcommand{\mc}[1]{\mathcal{#1}}

\newcommand{\va}{\mathbbm{a}}

\newif\ifabridged

\abridgedfalse

\begin{document}
	\title{Entanglement renormalization for quantum fields with boundaries and defects}
	\author{Adrián Franco-Rubio}
	\affiliation{Perimeter Institute for Theoretical Physics, Waterloo, ON, N2L 2Y5, Canada}
	\affiliation{Department of Physics and Astronomy, University of Waterloo, Waterloo, ON, N2L 3G1, Canada}
	\affiliation{Max-Planck-Institut für Quantenoptik, Hans-Kopfermann-Stra\ss e 1, 85748 Garching, Germany}
	\affiliation{Munich Center for Quantum Science and Technology, Schellingstra\ss e 4, 80799 München, Germany}
	
\begin{abstract}
	The continuous Multiscale Entanglement Renormalization Ansatz (cMERA) [Haegeman et al., Phys. Rev. Lett. 110, 100402 (2013)] gives a variational wavefunctional for ground states of quantum field theoretic Hamiltonians. A cMERA is defined as the result of applying to a reference unentangled state a unitary evolution generated by a \textit{quasilocal} operator, the \textit{entangler}. This makes the extension of the formalism to the case where boundaries and defects are present nontrivial. Here we show how this generalization works, using the 1+1d free boson cMERA as a proof-of-principle example, and restricting ourselves to conformal boundaries and defects. In our prescription, the presence of a boundary or defect induces a modification of the entangler localized only to its vicinity, in analogy with the so-called \textit{principle of minimal updates} for the lattice tensor network MERA.
\end{abstract}
	\maketitle
\section{Introduction}

\par The study and simulation of a quantum many-body system is generically a very challenging numerical problem, due to the exponential growth of the computational cost with the number of degrees of freedom. In order to facilitate this task, \textit{tensor network} ansätze for quantum states have been introduced. They make use of the particular entanglement structure of ground states of local Hamiltonians to provide an efficient way to represent and manipulate them (see \cite{Orus:2013kga,Bridgeman:2016dhh, tensorsnet, tensornetworkorg} for reviews and code examples). Among these, the multi-scale entanglement renormalization ansatz (MERA) has been proven successful at capturing relevant properties of ground states of critical Hamiltonians, such as the asymptotic scaling of their correlators and entanglement entropy \cite{Vidal:2007hda,Vidal:2008zz}. MERA has found application in areas ranging from topological order \cite{Aguado:2007oza} to error correction \cite{FerrisPoulin_2014}, machine learning \cite{beny2013deep} or quantum gravity \cite{Swingle_2012,Czech:2015kbp}.

\par In the last decade, \textit{continuous} tensor networks have arisen as analogous constructions to lattice tensor networks in the setting of quantum field theory (QFT) \cite{VerstraeteCirac, Haegeman:2011uy, Jennings:2015nwa, Tilloy:2018gvo, Hu:2018hyd}. The continuous MERA (cMERA) was introduced in \cite{Haegeman:2011uy} as an ansatz wavefunctional for the ground states of QFT Hamiltonians. cMERA states are defined by means of a unitary evolution generated by a \textit{quasilocal} operator, the \textit{entangler}, and, as a consequence, display a built-in UV cutoff length scale \cite{Franco-Rubio:2017tkt}. They have been proven capable of approximating the long distance properties of noninteracting QFT ground states with bosonic, fermionic and gauge degrees of freedom \cite{Franco-Rubio:2019nne}. In the particular case of conformal field theory (CFT) ground states, a full representation of the conformal group can be defined on the cMERA approximation, allowing for the extraction of conformal data \cite{Hu:2017rsp}. cMERA has also found applications in quantum gravity, as a toy model for the holographic principle and the AdS/CFT correspondence \cite{Nozaki:2012zj,Mollabashi:2013lya,Miyaji:2014mca,Molina-Vilaplana:2015mja, Chapman_2018}. 

\begin{figure}[t]
	\centering
	\includegraphics[width=1\linewidth]{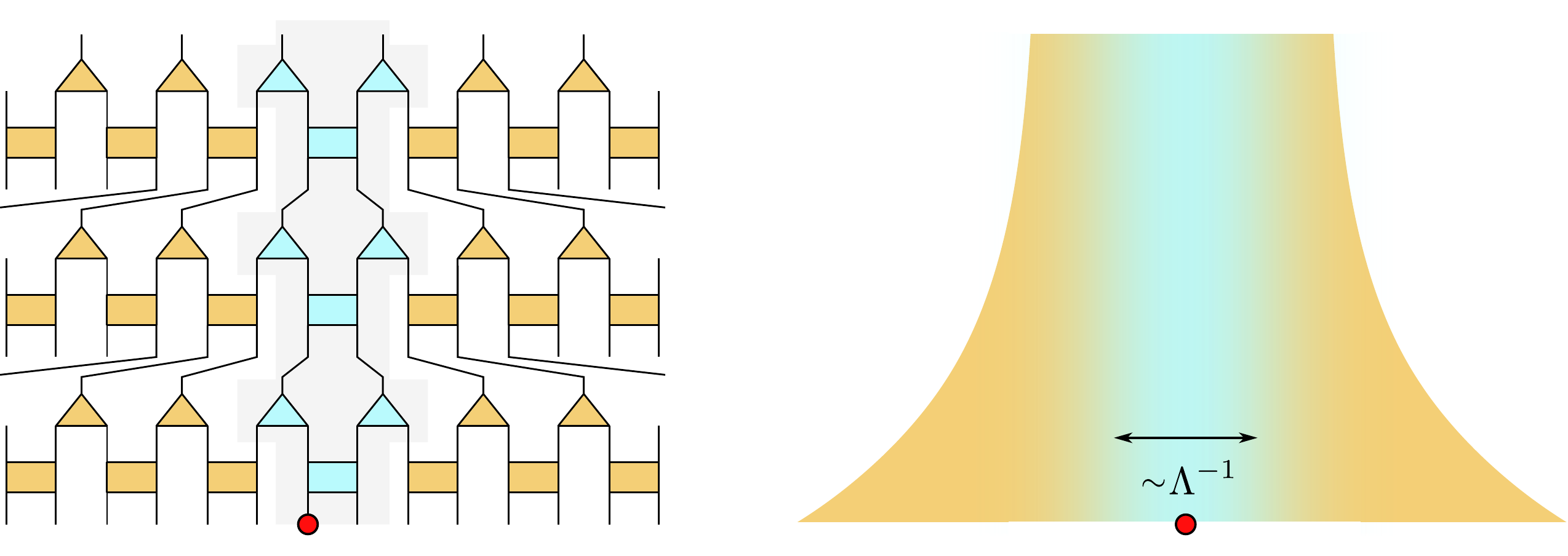}
	\caption{(left) The conjecture of minimal updates localizes the impact of a defect on the tensors of a MERA to its causal cone. (right) The effect of a defect in cMERA is analogously localized to a ``smeared causal cone'' in its vicinity.}
	\label{fig:causalcone}
\end{figure}

\par A QFT can be probed by placing it in a manifold with a boundary, or by introducing a defect (domain wall) separating two theories or two instances of the same theory. This intuitively requires a modification of the cMERA formalism, due to the finite characteristic length of the entangler. In this paper, we show explicitly how this modification works for the free scalar cMERA. This problem has strong ties to the situation on the lattice, where boundary critical phenomena were first addressed with entanglement renormalization techniques in \cite{Evenbly2010}. Later, Evenbly and Vidal analyzed MERA representations of systems with boundaries and impurities (defects) \cite{Evenbly2014}. They proposed and provided strong evidence for the \textit{minimal update conjecture} \cite{Evenbly:2013tta}, which states that the insertion of a boundary/defect only requires the modification of the tensors of the MERA within a localized region around it, called its \textit{causal cone} (see Fig.~\ref{fig:causalcone}, left). As a practical consequence, the state resulting from adding a boundary/defect to a translation invariant state can be parameterized with a smaller number of different tensors than a generic non-translation invariant state. Defects in MERA have recently received attention in the context of topological theories \cite{Hauru:2015abi, Bridgeman:2017etx} and holography \cite{Czech:2016nxc, Chapman:2018bqj}.

\par In this paper, we build cMERA evolutions for non-interacting scalar theories with boundaries and defects. These support an equivalent \textit{minimal update} prescription: that cMERA approximations for systems with a boundary or defect can be obtained from those without them by modifying the entangler within a ``smeared causal cone'', of width given by the quasilocality length scale (see Fig.~\ref{fig:causalcone}, right).
We expect this conjecture to hold indistinctly for free and interacting systems, as is the case for lattice MERA.

\par This paper is structured as follows: in Section \ref{sec:review} we provide a review of cMERA and the basics of boundaries and defects in the free boson CFT, in order to set up the notation for the following sections. In Section \ref{sec:cMERA} we present our main result: we build cMERAs for the free boson boundary and defect CFTs. We conclude with a discussion in Section \ref{sec:disc}.
	

\section{Review of relevant concepts}
	\label{sec:review}
	\subsection{Basics of cMERA}
		\label{subsec:cMERA}
		The cMERA contruction generates a one-parameter family $\ket{\Psi^\Lambda(s)}$ of ansatz wavefunctionals, which can be used to approximate the ground state of a given QFT \cite{Haegeman:2011uy}. They are defined by means of a unitary evolution applied to a reference state $\ket\Lambda$:
		\begin{equation}
			\ket{\Psi^\Lambda(s)} \equiv \mc P\exp\left(-i \int_0^s{du\;\left(L+K(u)\right)}\right)\ket\Lambda.
		\end{equation}
		Here $\mc P\exp$ stands for path-ordered integration and $L+K(u)$ is a Hermitian generator, which we explain below.
		
		\par Fundamentally, cMERA states should be understood as the result of an \textit{entangling evolution in scale}, i.e.\ the sequential introduction of entanglement in an initially unentangled state, in such a way that correlations are introduced progressively by length scale. To that end, we choose $\ket{\Lambda}$ to be 
		such that for local operators $\mc O, \mc O'$, their connected correlator vanishes at different points, i.e.,
		\begin{equation}
			\bra{\Lambda}\mc O (x)\mc O'(y)\ket\Lambda-\bra{\Lambda}\mc O (x)\ket\Lambda\bra{\Lambda}\mc O'(y)\ket\Lambda=0,
			\label{eq:unent}
		\end{equation}
		except possibly when $x=y$. This is equivalent to saying that $\ket{\Lambda}$ is the continuum limit of a product state on the lattice: it contains no entanglement.
		
		On the other hand, the Hermitian generator $L+K(u)$ is the sum of two contributions. $L$ is the generator of scale transformations, which rescales space and the fields, see e.g. Eqs. \eqref{eq:L1}-\eqref{eq:L2} below. By itself, $L$ cannot introduce entanglement in $\ket{\Lambda}$. That is the task for the \textit{entangler} $K(u)$, taken to be a quasilocal operator:  Eqs. \eqref{eq:free_K} and \eqref{profiles} show an important example. $K(u)$ entangles degrees of freedom around a fixed length scale, conventionally denoted $\Lambda^{-1}$ (so that $\Lambda$ is the associated momentum scale \footnote{Even though generally it need not be the case, in the free bosonic cMERAs we work with here, the initial state should be chosen depending on the entangling length scale $\Lambda^{-1}$, which justifies the notation $\ket{\Lambda}$.}). As the evolution progresses, $L$ keeps rescaling the correlations introduced by $K(u)$, so that $\ket{\Psi^\Lambda(s)}$ is entangled at length scales within $(\Lambda^{-1},\Lambda^{-1}e^s)$, while it remains practically uncorrelated at distances below $\Lambda^{-1}$. As a consequence, $\Lambda$ plays the role of a UV cutoff \cite{Franco-Rubio:2017tkt}.
		
		\par In the case where $K(u)\equiv K$ is $u$-independent (as in Eq. \eqref{eq:free_K}), there may exist an asymptotic fixed point cMERA state,
		\begin{equation}
			\ket{\Psi^\Lambda}\equiv\lim_{s\to\infty}{e^{-is(L+K)}}\ket{\Lambda}=\lim_{s\to\infty}\ket{\Psi^\Lambda(s)}.
		\end{equation}
		These are called \textit{scale invariant} cMERAs. With the exception of Appendix \ref{app:magic}, we will focus entirely on such fixed point cMERA states, which we use to approximate the ground states of conformally invariant target theories (CFTs, boundary CFTs and defect CFTs).
	
		
		\subsection{Example: free scalar cMERA}
		\label{subsec:cMERAexample}
		
		In this work, we will be making heavy use of the cMERA for the free scalar theory originally defined in \cite{Haegeman:2011uy}. Consider a bosonic field $\phi(x)$ defined on the real line, together with its conjugate momentum $\pi(x)$. The Klein-Gordon Hamiltonian
		\begin{equation}
			H = \ha\int{dx\;\left[(\pi(x))^2+(\partial\phi(x))^2\right]}
			\label{KleinGordon}
		\end{equation}
		has a unique ground state $\ket{\Psi}$ with two-point correlation functions
		\begin{align}
			\mc C_{\phi\phi}(x)&\equiv\bra{\Psi}\phi(0)\phi(x)\ket{\Psi}=-\dfrac{1}{2\pi}\log{|x|},
			\label{CFTcorrs1}\\
			\mc C_{\pi\pi}(x)&\equiv\bra{\Psi}\pi(0)\pi(x)\ket{\Psi}=-\dfrac{1}{2\pi}\dfrac{1}{x^2}.
			\label{CFTcorrs2}
		\end{align}
		Since $\ket\Psi$ is the ground state of a quadratic Hamiltonian, it belongs to the family of Gaussian states, and it is thus characterized by its two-point functions \footnote{The last two-point  function, $\bra{\Psi}\phi(0)\pi(x)\ket{\Psi}$ will equal $i/2\;\delta(x)$ for every Gaussian state in this work, and we will pay no further attention to it.}.
		\par
\begin{figure*}
		    \centering
		    \includegraphics[width=.95\textwidth]{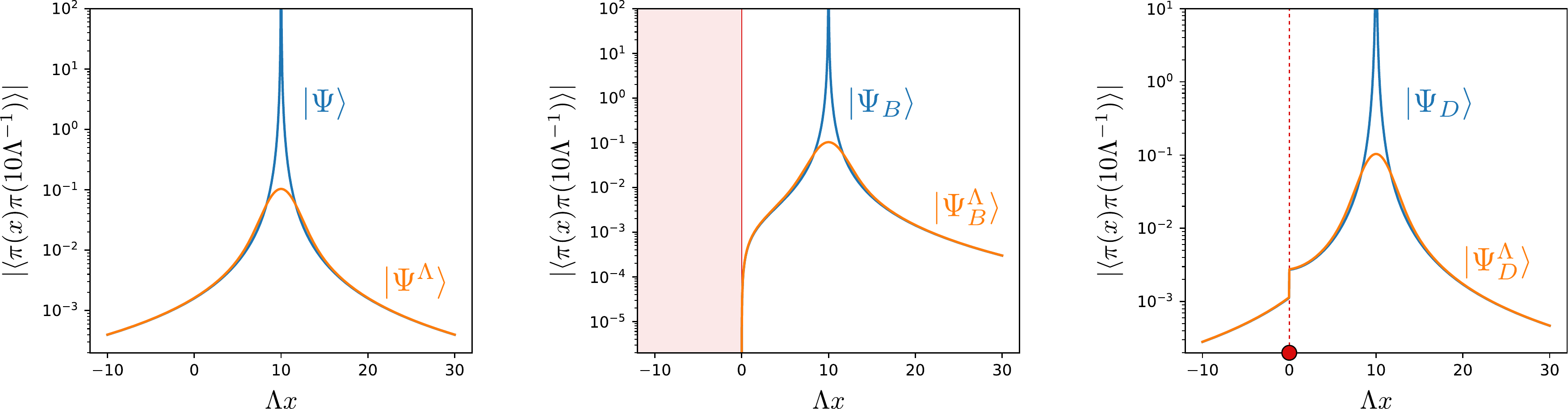}
		    \caption{The $\ev{\pi(x)\pi(y)}$ correlator with $y=10\Lambda^{-1}$ in the target ground state (blue) and its cMERA approximation (orange), for (left) the theory on the full line, (center) the theory on the half-line with Dirichlet b.c.\ at the origin, (right) the theory with a defect $\theta = \frac{3\pi}{8}$ at the origin. For each theory, its cMERA approximation reproduces the correlations accurately at distances $|x-y|\gg\Lambda^{-1}$. Delta function contributions are present at $x=y$ for the cMERA correlators but not depicted in the figure.}
		    \label{fig:corrs}
		\end{figure*}

 The cMERA approximation for $\ket\Psi$ is now given by specifying the three elements in its construction, namely $\ket\Lambda, L$ and $K$. These are chosen so that $\ket \Lambda$ is a Gaussian state, defined as the common kernel of local annihilation operators
		\begin{equation}
			\left(\sqrt{\dfrac{\Lambda}{2}}\phi(x)+i\sqrt{\dfrac{1}{2\Lambda}}\pi(x)\right)\ket\Lambda=0,\qquad \forall x,
			\label{unent2}
		\end{equation}
		(note that this implies that $\ket{\Lambda}$ satisfies condition \eqref{eq:unent}), and $L$ and $K$ are quadratic operators given by
		\begin{align}
			L &= \ha\int{dx\;\pi(x)\left(x\,\partial+\ha\right)\phi(x)}+\text{h.c.},\label{eq:freeL}\\
			K &= \int{dx\,dy\;g(x-y)\phi(x)\pi(y)}+\text{h.c.}\label{eq:free_K}
		\end{align}
		This all implies that the cMERA evolution takes place on the manifold of Gaussian states, which are easy to compute with. We therefore call it a Gaussian cMERA. The action of $L$ on the field operators is as follows
		\begin{align}
			 e^{isL}\phi(x)e^{-isL}&=e^{\ha s}\phi(e^sx), \label{eq:L1}\\
			 e^{isL}\pi(x)e^{-isL}&=e^{\ha s}\pi(e^sx), \label{eq:L2}
		\end{align}
		as befits a generator of scale transformations. On the other hand, the entangler $K$ is defined in terms of a function $g(x)$ which is taken to be quasilocal with characteristic length scale $\Lambda^{-1}$. Two examples of such a function are 
		\begin{equation}
			g(x) = \ha e^{-\frac{\sigma}{4}(\Lambda x)^2},\qquad g(x) = \dfrac{\Lambda}{4}e^{-\Lambda|x|}.
			\label{profiles}
		\end{equation}
		where $\sigma\equiv e^\gamma$ is the exponential of the Euler-Mascheroni constant. The former was essentially the choice made in \cite{Haegeman:2011uy}, while the latter corresponds to the \textit{magic} cMERA from \cite{Zou:2019xbi} (see also Appendix \ref{app:magic}). For an adequate choice of $g$ such as these, the fixed point state $\ket{\Psi^\Lambda}$ of this cMERA is a good long-distance approximation of $\ket\Psi$, in the sense that, for $x\gg\Lambda^{-1}$,
		\ifabridged
		\begin{eqnarray}
			\mc C^\Lambda_{\phi\phi}(x)&\equiv\bra{\Psi^\Lambda}\phi(0)\phi(x)\ket{\Psi^\Lambda}\approx\mc C_{\phi\phi}(x),
		\end{eqnarray}
		\else
	    \begin{eqnarray}
			\mc C^\Lambda_{\phi\phi}(x)&\equiv\bra{\Psi^\Lambda}\phi(0)\phi(x)\ket{\Psi^\Lambda}\approx\mc C_{\phi\phi}(x),
 			\\\mc C^\Lambda_{\pi\pi}(x)&\equiv\bra{\Psi^\Lambda}\pi(0)\pi(x)\ket{\Psi^\Lambda}\approx \mc         C_{\pi\pi}(x).
		\end{eqnarray}
		\fi
		that is, their correlation functions coincide at distances larger than than the UV cutoff scale $\Lambda^{-1}$. On the other hand, at length scales much shorter than $\Lambda^{-1}$, $C^\Lambda_{\phi\phi}(x)$ and $\mc C^\Lambda_{\pi\pi}(x)$ display a UV-regularized behaviour, reminiscent of $\ket\Lambda$, and tend to a constant in the $x\to 0$ limit (see Fig.~\ref{fig:corrs} (left)). The ultraviolet divergences from $\mc C_{\phi\phi}(x),\mc C_{\pi\pi}(x)$ are replaced by an on-site delta term such as the one in \eqref{eq:unent}  \cite{Franco-Rubio:2017tkt}.
		
		\subsection{Free bosonic theory: boundaries and defects}
		We summarize here the confomal boundary conditions and conformal defects of the free boson that we will use in what follows. For a more thorough review we refer the reader to Appendix \ref{app:boson}.
		\subsubsection{Conformal boundary conditions of the free boson}
		
		Consider once again a theory of a free scalar field, this time defined on the positive half-line $\R^+=\{x>0\}$. At $x=0$, we impose \textit{conformal} boundary conditions (b.c.). These turn the theory into a \textit{boundary} CFT (BCFT), i.e.\ one that is invariant under the subgroup of the conformal group that preserves the boundary.
		
		\par In our case, the following conformal b.c.\ exist:
		\begin{align}
			\phi(0)=0 & \qquad\text{(Dirichlet b.c.)}\\
			\partial_x\phi(0)=0 & \qquad\text{(Neumann b.c.)}
		\end{align}
		Each of these allows for the quantization of the theory to carry on in the standard fashion. The corresponding ground state $\ket{\Psi_B}$ is characterized by the following correlation functions, for $x,y>0$:
		\ifabridged
		\begin{align}
			\bra{\Psi_B}\phi(x)\phi(y)\ket{\Psi_B}=\mc C_{\phi\phi}(x-y) + \xi\,\mc C_{\phi\phi}(x+y),\label{eq:BCFTcorr1}
		\end{align}
		\else
		\begin{align}
			\bra{\Psi_B}\phi(x)\phi(y)\ket{\Psi_B}=\mc C_{\phi\phi}(x-y) + \xi\,\mc C_{\phi\phi}(x+y),\label{eq:BCFTcorr1}
			\\\bra{\Psi_B}\pi(x)\pi(y)\ket{\Psi_B}=\mc C_{\pi\pi}(x-y) + \xi\,\mc C_{\pi\pi}(x+y),
			\label{eq:BCFTcorr2}
		\end{align}
		\fi
		where $\xi=-1$ for Dirichlet boundary conditions, and $\xi = 1$ for Neumann boundary conditions. Note that translation invariance is lost, and the correlators satisfy the adequate b.c.
		
		\subsubsection{Conformal defects of the free boson}
		
		We use the word \textit{defect} to refer to a domain wall, sitting at the origin, that separates two instances of the free scalar theory, which respectively occupy the left and right half-lines $\R^-, \R^+$. Here we focus on \textit{conformal} defects, so that the resulting theory is a defect CFT (DCFT). In \cite{Bachas:2001vj}, two one-parameter families of conformal defects for the free boson are introduced. Of these, we will work with the first one \footnote{We explain this choice in Appendix \ref{app:boson}.}, for which the theories at each side of the defect are related by the matching conditions
		\begin{align}
			\partial_x\phi(0^-) &= \tan\theta\;\partial_x\phi(0^+),\\
			\partial_t\phi(0^-) &= \cot\theta\;\partial_t\phi(0^+),
		\end{align}
		where $\theta\in\left(-\frac{\pi}{2},\frac{\pi}{2}\right]$ is an angular parameter that characterizes the defect. Note that $\theta=\pi/4$ corresponds to the trivial defect (i.e.\ the absence thereof), while $\theta=0$ and $\theta=\pi/2$ correspond to pairs of boundary conditions, where the left and right theories completely decouple.
		
		\par Similarly to the BCFT, the resulting DCFT can be solved to obtain the two-point functions that characterize its Gaussian ground state $\ket{\Psi_D}$. For future reference, these can be written in terms of $\theta$ as follows
		\ifabridged
		\begin{align}
			\bra{\Psi_D}\phi(x)\phi(y)\ket{\Psi_D}=~&\mc C_{\phi\phi}(x-y)\nonumber\\&+c_\theta(x,y)\;\mc C_{\phi\phi}(|x|+|y|),
 			\label{hala2a}
		\end{align}
		\else
		\begin{align}
			\bra{\Psi_D}\phi(x)\phi(y)\ket{\Psi_D}=~&\mc C_{\phi\phi}(x-y)\nonumber\\&+c_\theta(x,y)\,\mc C_{\phi\phi}(|x|+|y|),
			\label{hala2a}\\
			\bra{\Psi_D}\pi(x)\pi(y)\ket{\Psi_D}=~&\mc C_{\pi\pi}(x-y)\nonumber\\&+c_\theta(x,y)\,\mc C_{\pi\pi}(|x|+|y|),
			\label{hala2b}
		\end{align}
        \fi
		by defining a piecewise constant function $c_\theta(x,y)$:
		\begin{equation}
			c_\theta(x,y)\equiv\begin{cases}
				\cos{2\theta} &  x,y<0,\\
				\sin{2\theta}-1 & xy<0,\\
				-\cos{2\theta} & x,y>0.
			\end{cases}
			\label{eq:piecewise_cte}
		\end{equation}
		

\section{cMERA in the presence of boundaries and defects}
\label{sec:cMERA}

\subsection{Boundary cMERA}
	Consider the free scalar BCFT on $\R^+$ with either Dirichlet or Neumann b.c.\ at the origin.	In this section, we propose a cMERA approximation $\ket{\Psi^\Lambda_B}$ to the BCFT ground state $\ket{\Psi_B}$, taking the cMERA for the full real line from Sect.\ \ref{subsec:cMERAexample} as a starting point. We do it by specifying the initial unentangled state $\ket{\Lambda_B}$, the scaling generator $L_B$ and the entangler $K_B$. After that we check that the fixed point of the corresponding cMERA evolution approximates $\ket{\Psi_B}$ at long distances.
	\par \textit{Initial state} As we did for $\ket{\Lambda}$ in Eq.~\eqref{unent2}, we define $\ket{\Lambda_B}$ in terms of local annihilation operators
	\begin{equation}
		\left(\sqrt{\dfrac{\Lambda}{2}}\phi(x)+i\sqrt{\dfrac{1}{2\Lambda}}\pi(x)\right)\ket{\Lambda_B}=0,\qquad\forall x>0.
	\end{equation}
	$\ket{\Lambda_B}$ can be interpreted as the continuum limit of a product state on a semi-infinite lattice.

	\textit{Scaling generator} We use the restriction to the half-line of the original generator $L$,
	\begin{equation}
		L_B \equiv L|_{\R^+} = \int_{x>0}{dx\;\pi(x)\left(x\,\partial+\ha\right)\phi(x)}+\text{h.c.}
	\end{equation}
    Note that $L_B$ preserves the half-line, and both b.c.

	\textit{Entangler} We define an entangler
	\begin{equation}
		K_{B} \equiv K|_{\R^+}+K_{\text{bdy}}.
	\end{equation}
	where $K|_{\R^+}$ is the restriction to the half-line of the original entangler $K$,
	\begin{equation}
		K|_{\R^+}\equiv\int_{x,y>0}{dx\,dy\;g(x-y)\phi(x)\pi(y)}+\text{h.c.},
	\end{equation}
	and the second term is
	\begin{equation}
		K_{\text{bdy}}\equiv\xi\int_{x,y>0}{dx\,dy\;g(x+y)\phi(x)\pi(y)}+\text{h.c.}.
	\end{equation}
	As before, $\xi=-1$ for Dirichlet b.c., and $\xi = 1$ for Neumann b.c. Due to the quasilocality of $g(x)$, this second term is supported mostly within a distance $\Lambda^{-1}$ of the boundary, and it leaves the original entangler nearly unchanged away from it. This property nicely corresponds with the minimal update conjecture from lattice MERA, in that the presence of the boundary only affects the entangling operator within a ``smeared'' causal cone of the boundary. Both in MERA and cMERA, because of the continuous rescaling as we evolve the system, the causal cone has a fixed width, of the order of the UV cutoff, i.e.\ the lattice spacing or $\Lambda^{-1}$, respectively.
	
In Appendix \ref{app:cMERA} we compute the correlators of the boundary cMERA to be
    \ifabridged
	\begin{align}
		\bra{\Psi^\Lambda_B}\phi(x)\phi(y)\ket{\Psi^\Lambda_B}=\mc C^\Lambda_{\phi\phi}(x-y) + \xi\,\mc C^\Lambda_{\phi\phi}(x+y),\label{eq:BcMERAcorrs1}
	\end{align}
	\else
		\begin{align}
		\bra{\Psi^\Lambda_B}\phi(x)\phi(y)\ket{\Psi^\Lambda_B}=\mc C^\Lambda_{\phi\phi}(x-y) + \xi\,\mc C^\Lambda_{\phi\phi}(x+y),\label{eq:BcMERAcorrs1}
		\\\bra{\Psi^\Lambda_B}\pi(x)\pi(y)\ket{\Psi^\Lambda_B}=\mc C^\Lambda_{\pi\pi}(x-y) + \xi\,\mc C^\Lambda_{\pi\pi}(x+y).
		\label{eq:BcMERAcorrs2}
	\end{align}
	\fi
From comparison with Eqs. \eqref{eq:BCFTcorr1}-\eqref{eq:BCFTcorr2}, it follows that the boundary cMERA will inherit the characteristic properties from the original cMERA: in particular, the correlation functions of $\ket{\Psi^\Lambda_B}$ approximate those of its target $\ket{\Psi_B}$ at distances longer than $\Lambda^{-1}$ (see Fig.~\ref{fig:corrs} (center), obtained using the first $g(x)$ from Eq. \eqref{profiles}). 

\subsection{Defect cMERA}
	Consider now the free scalar on the real line together with a defect insertion at the origin, parameterized by $\theta$ (we mostly omit the $\theta$ dependence in what follows in order not to overload the notation). Once again, we propose a cMERA construction $\ket{\Lambda_D}, L_D, K_D$, whose fixed point state $\ket{\Psi^\Lambda_D}$ approximates the long distance behavior of the ground state $\ket{\Psi_D}$ of the DCFT.
	
	\par \textit{Initial state and scaling generator} We choose them exactly as in the case without defect \footnote{It could be argued that there are no dynamical degrees of freedom at $x=0$, since the values of the field are constrained by the defect conditions. We will not worry about this since in any case we are speaking of a measure zero spatial set.}:
	\begin{equation}
		\ket{\Lambda_D}\equiv\ket{\Lambda},\qquad L_D \equiv L.
	\end{equation}
	Note that $L_D$ preserves the location of the defect at the origin.
	
	\textit{Entangler} We propose an entangling operator obtained from adding a modifying term to the defectless entangler 
	\begin{equation}
		K_{D} \equiv K+K_{\text{def}},
		\label{KD}
	\end{equation}
	where
	\begin{equation}
		K_{\text{def}}\equiv\dfrac{1}{2}\int_{x,y>0}{dx\,dy\;c_\theta(x,y)\,g(|x|+|y|)\,\phi(x)\pi(y)}+\text{h.c.}
		\label{Kdef}
	\end{equation}
	and $c_\theta(x,y)$ is given by \eqref{eq:piecewise_cte} \footnote{It can be seen that, for $\theta=0,\frac{\pi}{2}$, when the defect amounts to two independent boundary conditions, our defect cMERA prescription gives the same result as applying the boundary cMERA prescription to each independent boundary, i.e.\ the defect entangler breaks up into two uncorrelated boundary entanglers.}.
	As in the previous section, the quasilocality of $g$ implies that the modification due to $K_{\text{def}}$ nearly vanishes at distances larger than $\Lambda^{-1}$ from the defect,	resulting in the cMERA version of a minimal update, depicted in Fig.~\ref{fig:causalcone}. 
 	Furthermore, in Appendix \ref{app:cMERA} we prove
\ifabridged
	\begin{align}
		\bra{\Psi^\Lambda_D}\phi(x)\phi(y)\ket{\Psi^\Lambda_D}=~&\mc C^\Lambda_{\phi\phi}(x-y)\nonumber\\
		&+c_\theta(x,y)\,\mc C^\Lambda_{\phi\phi}(|x|+|y|),
	\end{align}
	\else
		\begin{align}
		\bra{\Psi^\Lambda_D}\phi(x)\phi(y)\ket{\Psi^\Lambda_D}=~&\mc C^\Lambda_{\phi\phi}(x-y)\nonumber\\
		&+c_\theta(x,y)\,\mc C^\Lambda_{\phi\phi}(|x|+|y|),
		\label{eq:DcMERAcorrs1}\\
		\bra{\Psi^\Lambda_D}\pi(x)\pi(y)\ket{\Psi^\Lambda_D}=~&\mc C^\Lambda_{\pi\pi}(x-y)\nonumber\\
		&+c_\theta(x,y)\,\mc C^\Lambda_{\pi\pi}(|x|+|y|).
		\label{eq:DcMERAcorrs2}
	\end{align}
\fi
	From the comparison to Eqs.~\eqref{hala2a}-\eqref{hala2b}, and from the properties of the cMERA correlators $\mc C^\Lambda_{\phi\phi}(x), \mc C^\Lambda_{\pi\pi}(x)$ it follows that our defect cMERA yields a good long distance approximation to the DCFT ground state, as exemplified by Fig.~\ref{fig:corrs} (right), obtained as well using the first $g(x)$ from Eq. \eqref{profiles}.


\section{Discussion and outlook}

\label{sec:disc}

In this paper we have introduced cMERA approximations for the ground states of BCFTs and DCFTs, reconciling the quasilocality of the entangler with the sharpness of boundaries and defects, and further expanding the application domain of the ansatz. For instance, a simple modification of our construction would allow for the definition of a cMERA on a compact interval with open boundary conditions, thus completing the formalism of \cite{Hung:2021tsu} where periodic boundary conditions are considered. Moreover, our results fit within the broader framework of the renormalization group approach to impurity problems, rooted in the Wilson's work \cite{RevModPhys.47.773}. In particular, we have shown that an analogue of the minimal update conjecture can be mathematically proved for noninteracting cMERAs. This complements the situation for lattice MERA, where heuristic evidence for the conjecture exists for both free and interacting systems. Together, these approaches hint at the role of minimal updates as a fundamental principle for the structure of quantum correlations in many-body systems.

It would be interesting to apply this formalism to the study of boundary critical phenomena. Following techniques from \cite{Hu:2017rsp}, it can be seen that fixed point boundary cMERA states support a full representation of the subgroup of conformal symmetries that characterize their target BCFT. This allows for the extraction of conformal data from the BCFT, analogously to the CFT case. On the other hand, defect cMERA provides a framework to study defect fusion rules. In critical lattice MERA, the fusion algebra of scaling operators can be extracted by coarse-graining operator insertions until their separation becomes smaller that the lattice spacing \cite{evenbly2013quantum}. Analogously, a backward cMERA evolution can bring two defects to within a distance of order $\Lambda^{-1}$, below which the cMERA stops resolving them, leading to defect fusion.

Our ability to carry out this example in full detail is undeniably linked to the fact that the theories used as proofs-of-principle are all non-interacting. This leads to the very clear parallel between the structure of the correlation functions and the entangler that generates them (e.g. between Eqs. \eqref{KD}-\eqref{Kdef} and \eqref{eq:DcMERAcorrs1}-\eqref{eq:DcMERAcorrs2}). We expect however that results for solvable systems such as ours will be able to provide useful prescriptions for interacting cMERA algorithms once these have reached their maturity (some approaches to the interacting case include \cite{Cotler:2018ehb,Cotler:2018ufx,Zou:2019xbi,Fernandez-Melgarejo:2019sjo,Fernandez-Melgarejo:2020fzw,numintercMERA}). In particular, we expand on the relation between our results and the proposal from \cite{Zou:2019xbi} in Appendix \ref{app:magic}.

\section*{Acknowledgements}
We are very grateful to Guifré Vidal, Qi Hu, Ignacio Cirac and Manuel Asorey for illuminating discussions. We would also like to thank the Instituto de Física Teórica in Madrid and X, the Moonshot Factory for hospitality during part of the completion of this work. Research at Perimeter Institute is supported by the Government of Canada through the Department of Innovation, Science and Economic Development Canada and by the Province of Ontario through the Ministry of Research, Innovation and Science.


\appendix
\section{Free boson BCFT and DCFT}
\label{app:boson}
In this appendix we review in more depth the BCFTs and DCFTs used in the main text.
\subsection{Free boson CFT}

Let us start by briefly recalling how to canonically quantize and solve the free boson theory on the real line, which is accomplished by expanding the field in momentum space. The action of the theory is given by
\begin{equation}
    S = \ha\int{dt\,dx\left[ (\partial_t\phi(x,t))^2-(\partial_x\phi(x,t))^2\right]}
\end{equation}
and the standard calculus of variations leads to the massless Klein-Gordon equation of motion,
\begin{equation}
    \left(\partial_t^2-\partial_x^2\right)\phi(x,t)=0,
    \label{eom}
\end{equation}
whose general solution reads
\begin{equation}
    \phi(x,t)=\int{\dfrac{dk}{\sqrt{4\pi|k|}}\; \left(a_k e^{i(kx-|k|t)}+a_k^\dagger e^{-i(kx-|k|t)}\right)},
    \label{modedecomp}
\end{equation}
The coefficients $a_k, a^\dagger_k$ can be written in terms of the momentum modes (Fourier components) of the fields
\begin{align}
    \phi(k)&\equiv \int{\dfrac{dx}{\sqrt{2\pi}}e^{-ikx}\phi(x)},\qquad \phi^\dagger(k)=\phi(-k),\\
    \pi(k)&\equiv\int{\dfrac{dx}{\sqrt{2\pi}}e^{-ikx}\pi(x)},\qquad \pi^\dagger(k)=\pi(-k),
\end{align}
as follows
\begin{equation}
    a_k=\sqrt{\dfrac{|k|}{2}}\phi(k)+i\sqrt{\dfrac{1}{2|k|}}\pi(k).
    \label{desire1}
\end{equation}
Now we promote the fields to operators and impose canonical commutation relations, which in our different sets of variables look, equivalently, like this 
\begin{align}
    [\phi(x),\pi(y)]&=i\delta(x-y),\label{comrel1}\\
    [\phi(k),\pi^\dagger(q)]&=i\delta(k-q),\label{comrel2}\\
    [a_k,a_q^\dagger]&=\delta(k-q).\label{comrel3}
\end{align}
The Klein-Gordon Hamiltonian can also be rewritten in momentum space, where all momenta decouple
\begin{align}
    H &= \ha\int{dx\,(\pi(x))^2+(\partial_x\phi(x))^2}\label{ham1}\\
      &= \ha\int{dk\,\left(\pi^\dagger(k)\pi(k)+k^2\phi^\dagger(k)\phi(k)\right)}\\
      &= \int{dk\; |k|a_k^\dagger a_k^{\phantom \dagger}}\label{ham3},
\end{align}
where the last equality is up to an infinite constant. The last expression allows us to easily characterize the vacuum of the theory as the common kernel of the annihilation operators
\begin{equation}
    a_k\ket{\Psi}=0,\qquad\forall k.
    \label{com1}
\end{equation}
The determination of the two-point functions of the vacuum, which completely determine it (due to it being a Gaussian state) follows from this characterization, yielding
\begin{align}
    \ev{\phi(k)\phi^\dagger(q)}&=\dfrac{1}{2|k|}\delta(k-q),\label{com2}\\
    \ev{\pi(k)\pi^\dagger(q)}&=\dfrac{|k|}{2}\delta(k-q).
    \label{com3}
\end{align}
By inverse Fourier transform, we get back the correlators $\mc C_{\phi\phi}(x), \mc C_{\pi\pi}(x)$ from Eqs. \eqref{CFTcorrs1}-\eqref{CFTcorrs2}.

\subsection{Free boson BCFT}

In order to define a quantum field theory on a spatial manifold with boundaries, it is important to establish the boundary conditions for the fields, in order to, among others, have a well-defined variational principle. Consider for instance the theory of a free scalar field on the half-line, given by the action
\begin{equation}
    S = \ha\int_{[0,T]\times\R^+}{dt\,dx\left[ (\partial_t\phi(x,t))^2-(\partial_x\phi(x,t))^2\right]}
\end{equation}
for a time interval $[0,T]$. The variation of the action given a variation of the field $\delta\phi(x,t)$ that vanishes at the endpoints of the time interval is then given by
\begin{align}
    \delta S &= \int_{[0,T]\times\R^+}{dt\,dx\left[ \partial_t\phi(x,t)\partial_t\delta\phi(x,t)\right.}\nonumber\\&~~~~~~~~~~~~~~~~~~~~~~~~~~~~~~\left.-\partial_x\phi(x,t)\partial_x\delta\phi(x,t)\right]\\
    &=\int_{[0,T]\times\R^+}{dt\,dx\;\left(-\partial^2_t\phi(x,t)+\partial^2_x\phi(x,t)\right)\delta\phi(x,t)}\nonumber\\&+\int_0^T{dt\;\partial_x\phi(0,t)\delta\phi(0,t)}.
\end{align}
If we now set $\delta S$ to 0 for an arbitrary $\delta\phi(x,t)$, apart from recovering the Klein-Gordon equation of motion \eqref{eom}, we are forced to impose $\partial_x\phi(0,t)=0$ (Neumann b.c.) or to fix the value of $\phi(0,t)$ so that it is no longer dynamical and $\delta\phi(0,t)=0$ (Dirichlet b.c.). Other approaches to the study of boundary condition include the functional analytic study of self-adjoint extensions of a symmetric Hamiltonian, and various charge conservation principles \cite{ReedSimon, Asorey:2004kk, Asorey:2015lja}.

Consider now as an example the Dirichlet case ${\phi(0,t)=0}$ (both this b.c.\ and the Neumann b.c.\ ${\partial_x\phi(0,t)=0}$ can be seen to be scale invariant, and thus will lead to boundary conformal field theories). Imposing the boundary condition in Eq. \eqref{modedecomp} restricts us to the subspace where $a_k=a_{-k}$, yielding
\begin{equation}
\phi(x,t)=\int_0^\infty{\dfrac{dk}{\sqrt{\pi |k|}}\;\left( a_k\sin{kx}\,e^{-i|k|t}+a_k^*\sin{kx}\,e^{i|k|t}\right)}.
\end{equation}
Note the domain of integration has been restricted to $k\geq 0$. If we modify the definition of $\phi(k),\pi(k)$ by using the \textit{sine}-Fourier transform
\begin{align}
\phi(k) &\equiv \sqrt{\dfrac{2}{\pi}}\int_{0}^\infty{dx\; \sin{kx}\,\phi(x)},\label{Dirichletmode1}\\\qquad \pi(k) &\equiv \sqrt{\dfrac{2}{\pi}}\int_{0}^\infty{dx\; \sin{kx}\,\pi(x)},\label{Dirichletmode2}
\end{align}
we have that Eq. \eqref{desire1} still holds, relating the new annihilation operators to the new momentum modes. So do the commutation relations \eqref{comrel1}-\eqref{comrel3} (note however that $\phi(k)$ and $\pi(k)$ are now Hermitian). Moreover, the Hamiltonian on the half-line can be represented analogously to the full line, i.e., Eqs. \eqref{ham1}-\eqref{ham3} hold upon restriction of the integrals to $x>0$ and $k>0$,
leading to a similar characterization of the ground state in terms of annihilation operators: Eqs. \eqref{com1}-\eqref{com3} hold as before, now restricted to positive values of $k$. Inverting the sine-Fourier transforms \eqref{Dirichletmode1}-\eqref{Dirichletmode2} we recover the Dirichlet version of \eqref{eq:BCFTcorr1}-\eqref{eq:BCFTcorr2}. The Neumann b.c.\ can be dealt with similarly (it basically amounts to replacing sines by cosines). 

As a side note, some insight into the BCFT correlators can be obtained by the \textit{method of images} (a.k.a. the \textit{doubling trick)} due to Cardy \cite{Cardy:1984bb}: notice that the new piece in the correlator of the BCFT can be interpreted as the usual two-point function for field insertions at $-x,y$ or $x,-y$, i.e.\ where one of the insertions has been reflected on the boundary.

\subsection{Free boson DCFT}

As presented in the main text, the notion of a defect we use is that of a domain wall between to (possible equal) theories. In this sense, boundaries are a particular kind of defect that separate a theory from the trivial theory (also sometimes called the vacuum theory, which has no degrees of freedom), and thus the boundary formalism is contained in the defect formalism. Here we focus on defects between two instances of the same theory, namely the free boson CFT. Whenever such a defect is conformal, i.e. scale invariant, we speak of a defect CFT (DCFT). This condition can be written as a condition on the continuity of the energy-momentum tensor of the CFT at the defect. Also in terms of the energy-momentum tensor, reflection and transmission coefficients $\mc R, \mc T$ (adding up to 1) can defined for a conformal defect. Defects for which $\mc R=1$ are \textit{totally reflective} or \textit{factorising}, and can be seen to correspond to two conformal boundary conditions, so that the theories at both sides of the defect are completely decoupled. On the other hand, defects for which $\mc T=1$ are \textit{totally transmissive} or \textit{topological}, the latter name stemming from the fact that they can be deformed arbitrarily without affecting correlation functions, as long as they do not cross field insertions. The trivial defect, the result of not really inserting anything and having a single theory on both sides, is an (equally trivial) example of such a defect.

In \cite{Bachas:2001vj}, two families of conformal defects for the free boson were introduced, each parameterized by an angular variable $\theta$. They are given by the matching conditions,
    \begin{align}
    	\partial_x\phi(0^-) &= \tan\theta\;\partial_x\phi(0^+),\label{match1}\\
    	\partial_t\phi(0^-) &= \cot\theta\;\partial_t\phi(0^+),\label{match1b}
    \end{align}
and	
	\begin{align}
		\partial_x\phi(0^-) &= \cot\theta\;\partial_t\phi(0^+),\label{match2}\\
		\partial_t\phi(0^-) &= \tan\theta\;\partial_x\phi(0^+).\label{match2b}
	\end{align}
The addition of these matching conditions, as in the boundary case, modifies the expansion of the field in momentum modes. Intuitively, the presence of the defect allows for the reflection of incoming plane waves, thus a wave travelling in a single direction will generally no longer be an eigenstate of the Hamiltonian. We are forced to consider a generalized momentum mode decomposition
\begin{equation}
    \phi(x,t)=\int{\dfrac{dk}{\sqrt{4\pi|k|}}\; \left(a_k f_k(x)e^{-i|k|t}+a_k^\dagger f^*_k(x)e^{i|k|t}\right)},
    \label{genmodedecomp}
\end{equation}
where the $f_k(x)$ are suitably chosen functions so that the equation of motion and the defect matching conditions are satisfied. We make the following ansatz 
\begin{align}
    f_k(x)=&\left(\alpha_k e^{ikx}+ \beta_k e^{-ikx}\right)\Theta(-x)\nonumber\\
    +&\left(\alpha'_k e^{ikx}+ \beta'_k e^{-ikx}\right)\Theta(x)
    \label{defectansatz}
\end{align}
where $\Theta(x)$ is the Heaviside step function. Clearly the resulting $\phi(x,t)$ satisfies the equation of motion at each side of the defect. For later reference, we compute the inner product of two such functions,
\begin{align}
    \langle f_k,f_q \rangle &=\int{dx\,f^*_k(x)f_q(x)}\\
    &=\left(\va_k^\dagger\va_q+{\va'_k}^\dagger\va'_q\right)\pi\delta(k-q)\nonumber\\
    &+\left(\va_k^\dagger X\va_q+{\va'_k}^\dagger X\va'_q\right)\pi\delta(k+q)\nonumber\\
    &+\left(\va_k^\dagger Z\va_q-{\va'_k}^\dagger Z\va'_q\right)\dfrac{i}{k-q}\nonumber\\
    &+\left(\va_k^\dagger ZX \va_q-{\va'_k}^\dagger ZX\va'_q\right)\dfrac{i}{k+q},
    \label{innerprod}
\end{align}
where we have defined
\begin{align}
    \va_k&\equiv\left(\begin{array}{c}
	\alpha_k\\
	\beta_k
	\end{array}\right), &\va'_k&\equiv\left(\begin{array}{c}
	\alpha'_k\\
	\beta'_k
	\end{array}\right),\\
    X&\equiv\left(\begin{array}{cc}
    0 & 1\\1 & 0
    \end{array}\right), &Z&\equiv\left(\begin{array}{cc}
    1 & 0\\0 & -1
    \end{array}\right),
\end{align}
and we have used the distributional expressions
\begin{align}
    \int_{-\infty}^0{dx\;e^{ikx}}=\pi\delta(k)-\dfrac{i}{k}\\
    \int_0^\infty{dx\;e^{ikx}}=\pi\delta(k)+\dfrac{i}{k}
\end{align}
Let us know impose the first set of matching conditions \eqref{match1}-\eqref{match1b}. If we define $\eta$ such that
\begin{equation}
    \cosh{\eta}\equiv\dfrac{\tan\theta+\cot\theta}{2},\qquad \sinh\eta\equiv\dfrac{\tan\theta-\cot\theta}{2},
\end{equation}
the defect conditions amount to
\begin{equation}
    	\va'_k=\left(\begin{array}{cc}
	\cosh\eta& \sinh\eta\\
	\sinh\eta & \cosh\eta
	\end{array}\right)\va_k\equiv R(\eta)\va_k,
	\label{gluingbasis}
\end{equation}
Thus, the matching conditions give the primed coefficients $\va'_k$ in \eqref{defectansatz} in terms of the unprimed ones $\va_k$. Finally, in order to recover the canonical commutation relations for the creation-annihilation operators, we look for a choice of $\alpha_k,\beta_k$ such that the $f_k$ form an orthonormal basis. Substituting \eqref{gluingbasis} in \eqref{innerprod}, the non-delta terms automatically vanish \footnote{To simplify the reasoning, notice that $R(\eta)=\exp(\eta X)$ and thus $[X,R(\eta)]=0$ and $R^\dagger(\eta)ZR(\eta)=Z$.}, so that solutions for momenta of different magnitude are orthogonal, and we are left with
\begin{align}
    \langle f_k,f_q \rangle &=\va_k^\dagger\left(\mathds{1}+R(\eta)^2\right)\va_q\,\pi\delta(k-q)\\ &+\va_k^\dagger\left(X+R(\eta)XR(\eta)\right)\va_q\,\pi\delta(k+q).
\end{align}
To arrive at an orthonormal basis, we can choose $\va_k\equiv\va$ to be independent of $k$, and such that 
\begin{align}
    \va^\dagger\pi\left(\mathds{1}+R^2(\eta)\right)\va&=1,\\
    \va^\dagger\left(X+R(\eta)XR(\eta)\right)\va&=0.
\end{align}
This can be achieved by introducing $\mathbbm{b}$ defined as ${\mathbbm{b}\equiv\sqrt{\pi}\left(\mathds{1}-iR(\eta)\right)\va}$, so that
\begin{align}
    \mathbbm{b}^\dagger\mathbbm{b}&=1,\\
    \mathbbm{b}^\dagger X\mathbbm{b}&=0.
\end{align}
Thus $\mathbbm{b}\in\{(1,0)^T, (0,1)^T\}$ up to a phase. We make the choice $\mathbbm{b}=(e^{i\frac{\pi}{4}},0)^T$, and revert all changes of variables to get our basis of functions
\begin{align}
    f_k(x)\equiv&\dfrac{e^{i\frac{\pi}{4}}}{2\sqrt{\pi}}\left[\left((\sin{2\theta}-i) e^{ikx} -i\cos{2\theta} e^{-ikx}\right)\Theta(-x)\right.\nonumber\\
    +&\left.\left((1-i\sin{2\theta}) e^{ikx} -\cos{2\theta} e^{-ikx}\right)\Theta(x)\right].
\end{align}
This choice for $\mathbbm{b}$ is motivated by the fact that, for $\theta=\frac{\pi}{4}$ (the trivial defect) we recover the plane wave basis. Now we can proceed to define our generalized momentum space fields:
\begin{equation}
    \phi(k)\equiv\int{dx\;f^*_k(x)\phi(x)},\qquad \pi(k)\equiv\int{dx\;f^*_k(x)\pi(x)}.
    \label{eq:defectmomfields}
\end{equation}
and once more, Eqs. \eqref{desire1}-\eqref{com3} hold. (Note however that Hermitian conjugation looks a bit different,
\begin{equation}
    \phi^\dagger(k) = \sin{2\theta}\phi(-k)-i\cos{2\theta}\phi(k)
\end{equation}
due to the presence of the defect.) Applying the inverse transformation, we get the correlators in real space, for instance:
\begin{align}
    \ev{\phi(x)\phi(y)} &= \int{dk\,dq\;f_k(x)f_q(y)\ev{\phi(k)\phi(q)}},\\
    &= \int{\dfrac{dk}{2|k|}f_k(x)f_k(y)}.
\end{align}
Solving the integrals leads back to Eqs. \eqref{hala2a}-\eqref{hala2b}. The transmission and reflection coefficients for these defects are given by
\begin{equation}
    \mc R = \cos^2{2\theta},\qquad \mc T = \sin^2{2\theta},
\end{equation}
which identifies the totally transmissive cases as $\theta = \frac{\pi}{4}$ (the trivial defect) and $\theta = -\frac{\pi}{4}$ (the $\pi$-phase defect, which changes the sign of the field when crossing it), and the totally reflective cases as $\theta=0, \frac{\pi}{2}$, which were already identified in the main text as the Neumann-Dirichlet and Dirichlet-Neumann pairs of boundary conditions.

If we now attempt the same strategy with the second family of defects (Eqs. \eqref{match2}-\eqref{match2b}), we run into trouble. The same ansatz leads to defect conditions
\begin{equation}
    	\va'_k=\left(\begin{array}{cc}
	\cosh\eta& -\sinh\eta\\
	\sinh\eta & -\cosh\eta
	\end{array}\right)\va_k\equiv -R(\eta)Z\va_k,
	\label{gluingbasis2}
\end{equation}
which then give rise to the following inner product
\begin{align}
    \langle f_k,f_q \rangle &=\va_k^\dagger\left(\mathds{1}+R(-\eta)^2\right)\va_q\,\pi\delta(k-q)\\ &+\va_k^\dagger\left(X-R(-\eta)XR(-\eta)\right)\va_q\,\pi\delta(k+q)\\ &+2\va_k^\dagger ZX\va_q\,\dfrac{i}{k+q}.
\end{align}
The last line implies that the subspaces of solutions with different momenta $k,q$ are generically not orthogonal, leading to nonvanishing overlaps between solutions with different momenta. This prevents us from writing a decomposition in canonically commuting creation-annihilation modes such as \eqref{genmodedecomp}, so the formalism we have been developing, where everything decouples in momentum space, does not apply. We will thus not deal with this family of defects any further. Note there are two exceptions to this behaviour, for $\theta=0,\pi$. These correspond to totally reflective defects, and give us the remaining choices of pairs of boundary conditions not included in the first family: Dirichlet-Dirichlet for $\theta = 0$ and Neumann-Neumann for $\theta=\pi$.


\section{cMERA, BcMERA and DcMERA}
\label{app:cMERA}
In this appendix we perform the computations that prove the properties we claim in the main text for the proposed boundary cMERA (BcMERA) and defect cMERA (DcMERA). We will progress in parallel to Appendix \ref{app:boson}, starting from the CFT, introducing boundary conditions, then defects.
\subsection{cMERA}
Let us first review the situation for the cMERA for free boson CFT, introduced in \cite{Haegeman:2011uy}.
We also use the formalism and notation from \cite{Hu:2017rsp}. The key to being able to solve the cMERA evolution exactly lies in the fact that we are dealing with translation invariant Gaussian states, which are characterized as common kernels of families of momentum-indexed annihilation operators. For instance, the CFT ground state $\ket{\Psi}$:
\begin{equation}
    a_k\ket{\Psi}=\left(\sqrt{\dfrac{|k|}{2}}\phi(k)+i\sqrt{\dfrac{1}{2|k|}}\pi(k)\right)\ket{\Psi}=0,
    \label{targetalpha}
\end{equation}
or the initial state $\ket{\Lambda}$
\begin{equation}
    \left(\sqrt{\dfrac{\Lambda}{2}}\phi(k)+i\sqrt{\dfrac{1}{2\Lambda}}\pi(k)\right)\ket{\Lambda}=0,
    \label{alphaesLambda}
\end{equation}
as one obtains from Fourier transforming Eq. \eqref{unent2}. All states $\ket{\Psi^\Lambda(s)}$ of the cMERA evolution can be written in this form, and thus parameterized by a single function $\alpha(k,s)$:
\begin{equation}
     \left(\sqrt{\dfrac{\alpha(k,s)}{2}}\phi(k)+i\sqrt{\dfrac{1}{2\alpha(k,s)}}\pi(k)\right)\ket{\Psi^\Lambda(s)}=0.
     \label{eq:cMERAannihilators}
\end{equation}
Indeed, we can obtain the annihilation operators for $\ket{\Psi^\Lambda(s)}=e^{-is(L+K)}\ket{\Lambda}$ by unitarily evolving those of the initial state $\ket\Lambda$:
\begin{equation}
    a\ket{\Lambda}=0\implies e^{-is(L+K)}ae^{is(L+K)}\ket{\Psi^\Lambda(s)}=0.
    \label{evolve_annihilators}
\end{equation}
In this way, we can derive the differential equation for $\alpha(k,s)$ corresponding to the cMERA evolution. This involves writing the generators $L$ and $K$ from Eqs. \eqref{eq:freeL}-\eqref{eq:free_K} in momentum space,
\begin{align}
    L&=\int{dk\;\pi^\dagger(k)\left(k\partial_k+\dfrac{1}{2}\right)\phi(k)}+\text{h.c.},\label{eq:LandKinmomentum1}\\
    K&=\ha\int{dk\;g(k)\phi(k)\pi^\dagger(k)}+\text{h.c.},
    \label{eq:LandKinmomentum2}
\end{align}
defining their action on the momentum fields $\phi(k),\pi(k)$,
\begin{align}
    \phi(k,s)&\equiv e^{i(L+K)s}\phi(k)e^{-i(L+K)s},\\
    \pi(k,s)&\equiv e^{i(L+K)s}\pi(k)e^{-i(L+K)s}.
\end{align}
and computing its infinitesimal form,
\begin{align}
\partial_s\phi(k,s)&=-i[L+K,\phi(k,s)]\nonumber\\&=-\left(k\partial_k+\dfrac{1}{2}+g(k)\right)\phi(k,s),\\
\partial_s\pi(k,s)&=-i[L+K,\pi(k,s)]\nonumber\\&=-\left(k\partial_k+\dfrac{1}{2}-g(k)\right)\pi(k,s),
\end{align}
where $g(k)$ is the Fourier transform of $g(x)$. Using these expressions together with \eqref{eq:cMERAannihilators} and \eqref{evolve_annihilators}, a differential equation for $\alpha(k,s)$ can be found:
\begin{equation}
    \partial_s\alpha(k,s) = \left(k\partial_k - 2g(k)\right)\alpha(k,s).
\end{equation}
The solution to this equation reads
\begin{equation}
    \alpha(k,s)=\alpha(ke^s,0)\exp{\left(-2\int_0^s{du\;g(ke^u)}\right)},
\end{equation}
and adding in the initial condition $\alpha(k,0)=\Lambda$ (from Eq. \eqref{alphaesLambda}),
\begin{equation}
    \alpha(k,s)=\Lambda\exp{\left(-2\int_0^s{du\;g(ke^u)}\right)}.
    \label{eq:sol_pde}
\end{equation}
This characterizes any cMERA state $\ket{\Psi^\Lambda(s)}$, in particular the asymptotic fixed point state $\ket{\Psi^\Lambda}$, given by 
\begin{equation}
    \alpha(k)\equiv\lim_{s\to\infty}{\alpha(k,s)}=\Lambda\exp{\left(-2\int_0^\infty{du\;g(ke^u)}\right)}.
    \label{eq:fixedpointalpha}
\end{equation}
For example, the two examples of $g(k)$ given in Eq. \eqref{profiles} lead, respectively, to 
\begin{align}
    \alpha(k) &= \Lambda \exp{\left(\ha \text{Ei}\left(-\dfrac{k^2}{\sigma\Lambda^2}\right)\right)},\\
    \alpha(k) &= \dfrac{\Lambda|k|}{\sqrt{k^2+\Lambda^2}},
\end{align}
where $\text{Ei(x)}$ is the exponential integral function. At the level of $\alpha$ functions the distinction between the short and long distance behaviour of the cMERA is even more transparent. For the two entanglers, it interpolates between $\alpha(k)\sim|k|$ at low momenta (long distances), which is the behaviour of the target ground state (see Eq. \eqref{targetalpha}) and $\alpha(k)\sim\Lambda$ at high momenta (short distances), which is the behaviour of the initial state (see Eq. \eqref{alphaesLambda}). Here ``low'' and ``high'' momenta are of course taken with respect to the cutoff $\Lambda$.

The two-point functions in the cMERA state can also be expressed rather compactly in terms of $\alpha(k)$:
\begin{align}
    \bra{\Psi^\Lambda}\phi(k)\phi^\dagger(q)\ket{\Psi^\Lambda}&=\dfrac{1}{2\alpha(k)}\delta(k-q),\label{alphacorr1}\\
    \bra{\Psi^\Lambda}\pi(k)\pi^\dagger(q)\ket{\Psi^\Lambda}&=\dfrac{\alpha(k)}{2}\delta(k-q).
    \label{alphacorr2}
\end{align}
Because of this, the comment we just made about the cMERA $\alpha$ function approximating the target ground state at low momenta/long distances translates into the correlation functions $\mc C_{\phi\phi}(x),C_{\phi\phi}(x)$, which is the statement we made in the main text.

\subsection{BcMERA}

Since our BcMERA is Gaussian, we enjoy a similar simple characterization of BcMERA states in terms of annihilation operators
\begin{equation}
\left(\sqrt{\dfrac{\alpha(k,s)}{2}}\phi(k)+i\sqrt{\dfrac{1}{2\alpha(k,s)}}\pi(k)\right)\ket{\Psi_B^\Lambda(s)}=0
\end{equation}
where the momentum fields $\phi(k),\pi(k)$ are now the ones defined in Eqs. \eqref{Dirichletmode1}-\eqref{Dirichletmode2} (resp. their Neumann version). Remember that in this BCFT context, $\phi(k),\pi(q)$ are Hermitian, and only defined for $k\geq0$.
The key to understanding our proposal for the BcMERA comes from expressing $L_B$ and $K_B$ as well in terms of these fields. If we do so, we notice their are again given by Eqs. \eqref{eq:LandKinmomentum1}-\eqref{eq:LandKinmomentum2}. Consequently, the derivation from the previous section follows through in the same fashion, reaching the analogue of Eqs. \eqref{alphacorr1}-\eqref{alphacorr2},
\begin{align}
    \bra{\Psi_B^\Lambda}\phi(k)\phi^\dagger(q)\ket{\Psi_B^\Lambda}&=\dfrac{1}{2\alpha(k)}\delta(k-q),\label{alphaBcorr1}\\ \bra{\Psi_B^\Lambda}\pi(k)\pi^\dagger(q)\ket{\Psi_B^\Lambda}&=\dfrac{\alpha(k)}{2}\delta(k-q),
    \label{alphaBcorr2}
\end{align}
where $\alpha(k)$ is again given \eqref{eq:fixedpointalpha}. Undoing the sine (resp. cosine)-Fourier transforms in Eqs. \eqref{Dirichletmode1}-\eqref{Dirichletmode2} (resp. their Neumann version), we prove Eqs. \eqref{eq:BcMERAcorrs1}-\eqref{eq:BcMERAcorrs2}.

\subsection{DcMERA}

It will not surprise the reader that the strategy we follow to propose the DcMERA is entirely analogous to the one we just went through for the BcMERA. All DcMERA states will be parameterized by an $\alpha$ function that characterizes their annihilation operators
\begin{equation}
\left(\sqrt{\dfrac{\alpha(k,s)}{2}}\phi(k)+i\sqrt{\dfrac{1}{2\alpha(k,s)}}\pi(k)\right)\ket{\Psi_D^\Lambda(s)}=0
\end{equation}
where the momentum fields $\phi(k),\pi(k)$ are the ones defined in Eq. \eqref{eq:defectmomfields}. The generators $L_D$ and $K_D$, written in terms of this DCFT momentum fields are again given by Eqs. \eqref{eq:LandKinmomentum1}-\eqref{eq:LandKinmomentum2}, and repeating our computations we reach
\begin{align}
    \bra{\Psi_D^\Lambda}\phi(k)\phi^\dagger(q)\ket{\Psi_D^\Lambda}&=\dfrac{1}{2\alpha(k)}\delta(k-q),\label{alphaDcorr1}\\ \bra{\Psi_D^\Lambda}\pi(k)\pi^\dagger(q)\ket{\Psi_D^\Lambda}&=\dfrac{\alpha(k)}{2}\delta(k-q),
    \label{alphaDcorr2}
\end{align}
which, undoing the transformation \eqref{eq:defectmomfields}, are equivalent to Eqs. \eqref{eq:DcMERAcorrs1}-\eqref{eq:DcMERAcorrs2}.


\section{Magic BcMERA and DcMERA}
\label{app:magic}
So far, our understanding of interacting cMERA is rather limited. One of the proposed roads towards the development of interacting cMERA algorithms is based on the \textit{magic} cMERA introduced in \cite{Zou:2019xbi}. In this Appendix, we briefly review what the magic cMERA is and then analyze the properties of its boundary and defect analogues.

\subsection{Magic cMERA}
We have, in fact, already shown in the main text the original example of a magic cMERA, namely the free boson cMERA whose entangler has the second profile from Eq. \eqref{profiles}:
\begin{equation}
    g(x) = \dfrac{\Lambda}{4}e^{-\Lambda|x|}.
    \label{eq:magic_g}
\end{equation}
This choice grants the construction the following nongeneric properties:
\par \textit{Local parent Hamiltonians} Each state $\ket{\Psi^\Lambda(s)}$ in the cMERA evolution, including the fixed point $\ket{\Psi}$, has a local parent Hamiltonian, i.e.\ it is the exact ground state of an operator that is the integral of a local density (generically for a cMERA state, it would only be quasilocal). Moreover, this Hamiltonian is obtained by adding to the CFT Hamiltonian $H$ from Eq. \eqref{KleinGordon} a UV cutoff term and an $s$-dependent IR cutoff term, which is nothing but a standard mass term:
    \begin{align}
        H^\Lambda(s) = H &+\dfrac{1}{2\Lambda^2}\int{dx\;(\partial_x\pi(x))^2}\\&+ \dfrac{1}{2}\int{dx\;m(s)^2\phi(x)^2},
        \label{eq:magicHamiltonian}
    \end{align}
where $m(s)\equiv\Lambda e^{-s}$. Notice how in this case evolving forward in $s$ corresponds very clearly to the lifting of the IR cutoff. Recall that, as mentioned in the introduction to cMERA, $\ket{\Psi^\Lambda(s)}$ presents correlations at length scales between those associated to the two cutoffs, $\Lambda^{-1}$ and $m(s)^{-1}=\Lambda^{-1}e^s$.

The Hamiltonian \eqref{eq:magicHamiltonian} can be found by the following reasoning in momentum space. For any given $s$, consider the family of Hamiltonians, 
\begin{equation}
    H[\varepsilon(k,s)] \equiv \int{dk\;\varepsilon(k,s)\,{a^\Lambda}^\dagger(k,s){a^\Lambda}(k,s)},
\end{equation}
where $\varepsilon(k,s)$ is a real positive function, and $a^\Lambda(k,s)$ the annihilation operators of the cMERA state $\ket{\Psi^\Lambda(s)}$ (cf. Eq. \eqref{eq:cMERAannihilators}):
\begin{equation}
    a^\Lambda(k,s) \equiv \sqrt{\dfrac{\alpha(k,s)}{2}}\phi(k)+i\sqrt{\dfrac{1}{2\alpha(k,s)}}\pi(k).
    \label{eq:cMERA_anni_expl}
\end{equation}
For the magic $g(k)$ from \eqref{eq:magic_g}, \eqref{eq:sol_pde} yields
\begin{equation}
    \alpha(k,s)=\dfrac{\sqrt{k^2+m^2(s)}}{k^2+\Lambda^2}.
\end{equation}
Note how, by construction, the cMERA state $\ket{\Psi^\Lambda(s)}$, which is characterized by 
\begin{equation}
    a^\Lambda(k,s)\ket{\Psi^\Lambda(s)}=0,
\end{equation}
is the ground state of $H[\varepsilon(k,s)]$ for any $\varepsilon(k,s)$. In fact, all Hamiltonians $H[\varepsilon(k,s)]$ have the same eigenvectors, while they differ on their spectrum (determined by the dispersion relation $\varepsilon(k,s)$). We can then choose the latter so that the Hamiltonian is local. It can be seen that the choice 
\begin{equation}
    \varepsilon(k,s)\equiv\dfrac{\sqrt{k^2+m^2(s)}\sqrt{k^2+\Lambda^2}}{\Lambda},
    \label{eq:magic_disp_rel}
\end{equation}
yields the local parent Hamiltonian $H^\Lambda(s)$ from above. 
\par
\begin{figure*}
    \centering
    \includegraphics[width=0.95\textwidth]{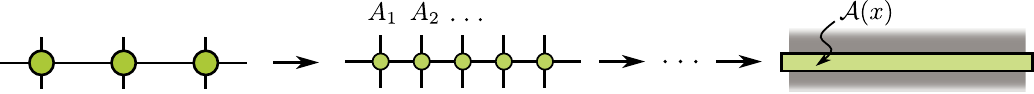}
    \caption{The continuum limit of discrete matrix product operators (MPO) gives rise to a continuous matrix product operator (cMPO). Both are characterized by finite-dimensional matrices $A_i$ (resp $\mc A(x)$), indexed by a discrete (resp. continuous) position label, whose entries are operators acting on the corresponding lattice site $i$ (resp. point $x$ on the line).}
    \label{fig:cMPO}
\end{figure*}

 \textit{Compatibility with cMPS and cMPO} Continuous matrix product states (cMPS) \cite{VerstraeteCirac} are probably the best understood family of continuous tensor network states. They are obtained from a well-defined continuum limit of lattice matrix product states (MPS). In a magic cMERA, all states $\ket{\Psi^\Lambda(s)}$ have the UV structure of a cMPS, and thus can be represented within this variational class. Moreover, the magic entangler $K$ can be represented as a continuous matrix product operator (cMPO, see Fig. \ref{fig:cMPO}), the continuous limit of a matrix product operator (MPO). The compatibility of the cMERA and cMPS/cMPO formalisms is crucial for the study of the interacting case, since cMPS techniques work for both Gaussian and non-Gaussian states and can thus handle the strongly correlated wavefunctionals required to represent interacting QFT ground states.

In \cite{Zou:2019xbi}, the fact that the UV structure of magic cMERA states is compatible with cMPS is argued in a dual manner. On the one hand, the local parent Hamiltonians $H^\Lambda(s)$ can be rewritten in terms of the creation-annihilation operators for the unentangled state $\ket{\Lambda}$ (cf. Eq. \eqref{alphaesLambda}), which we denote
    \begin{equation}
        \psi(x)\equiv\sqrt{\dfrac{\Lambda}{2}}\phi(x)+i\sqrt{\dfrac{1}{2\Lambda}}\pi(x),
        \label{eq:annimagic}
    \end{equation}
to yield
\begin{align}
 H^{\Lambda}_m = \int dx  &\left( \frac{1}{\Lambda} \partial_x \psi^{\dagger}(x)\partial_x \psi(x) + \frac{\Lambda^2 + m^2}{2\Lambda}\psi^{\dagger}(x)\psi(x)\right. \nonumber \\       
 &-\left. \frac{\Lambda^2 - m^2}{4\Lambda} \left(\psi(x)^2 + \psi(x)^{\dagger 2}\right) \right),
\label{eq:HLam}
\end{align}
whose first term, dominant at high energies, is a non-relativistic kinetic term, characteristic of Hamiltonians whose ground state is well approximated by cMPS. On the other hand, one can look at the occupation number with respect to the same creation-annihilation operators, $n(k,s)$, defined by
\begin{equation}
    \bra{\Psi^\Lambda(s)}\psi^\dagger(k)\psi(q)\ket{\Psi^\Lambda(s)}\equiv n(k,s)\delta(k-q).
\end{equation}
Its scaling at high energies $|k|\gg\Lambda$ can be computed to be
\begin{equation}
    n(k,s) = \dfrac{1}{4}\left(\dfrac{\alpha(k,s)}{\Lambda}+\dfrac{\Lambda}{\alpha(k,s)}-2\right)\sim \dfrac{1}{k^4}
    \label{eq:ndek}
\end{equation}
which is known to be the scaling of this quantity for a generic cMPS \cite{Haegeman:2010fv}.

\begin{figure*}
    \centering
    \includegraphics[width=.95\textwidth]{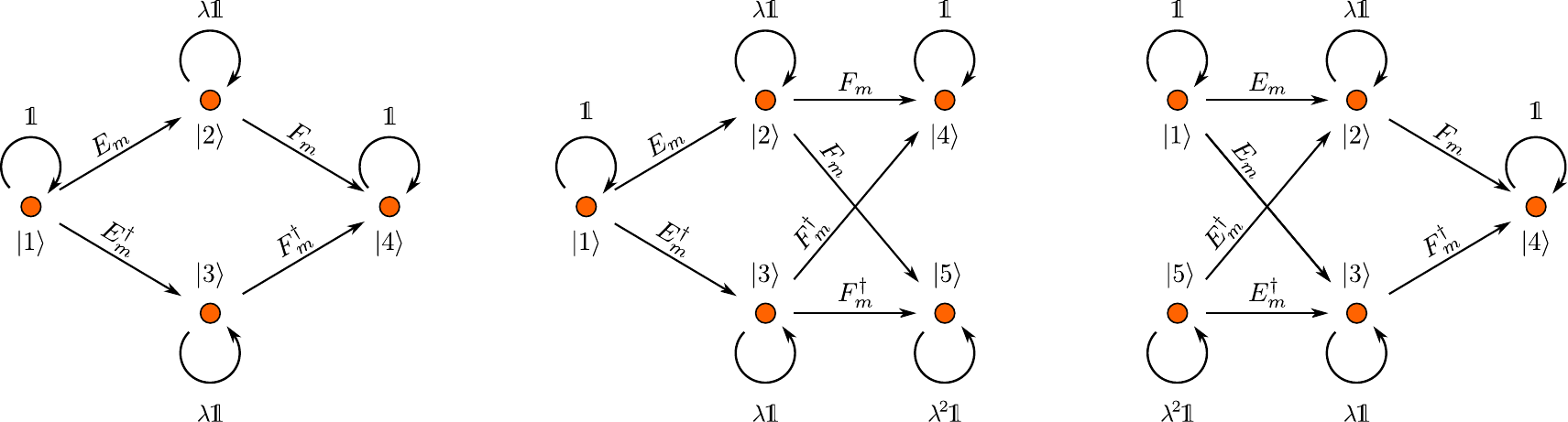}
    \caption{Finite state automata representations of (left) the MPO precursor of the magic cMERA cMPO $\mc A_K(x)$ from \eqref{eq:CFTcMPO}, (center) the MPO precursor of $\mc A_L(x)$ from \eqref{bdyAL}, (right) the MPO precursor of $\mc A_R(x)$ from \eqref{bdyAR}. The definitions of $E_m, F_m$ and $\lambda$ are as in \eqref{def_E}-\eqref{def_l}.}
    \label{fig:automata}
\end{figure*}

Let us now be more explicit about the expression of $K$ as a cMPO, following closely the Appendices from \cite{Zou:2019xbi}. In terms of $\psi(x),\psi^\dagger(x)$, the magic entangler reads
    \begin{equation}
        K=\dfrac{-i\Lambda}{8}\int{dx\,dy\;e^{-|x-y|}\psi(x)\psi(y)}+\text{h.c.}
    \end{equation}
cMPOs are defined as the continuum limit of some ``precursor'' MPO (see Fig. \ref{fig:cMPO}). Recall that an MPO is defined in terms of a series of matrices $A_m$ on a finite dimensional \textit{virtual} Hilbert space, whose entries are themselves operators acting on the \textit{physical} Hilbert space of the corresponding lattice site $m$. The dimensionality $\chi$ of the virtual Hilbert space (i.e. the size of the $A_m$) is the \textit{bond dimension} of the MPO, and we label its basis elements by $\ket{i}, i=1,\ldots,\chi$. The full operator the MPO represents is obtained as a virtual space matrix element:
\begin{equation}
    \mc O_{\text{MPO}} \equiv \bra{v}A_1A_2\ldots A_N\ket{w},
\end{equation}
where $\bra{v},\ket{w}$ are the boundary vectors in the extremes of the MPO, which tell us what matrix element to pick \footnote{Sometimes, periodic boundary conditions are imposed instead, which would amount to replacing the matrix element by a trace.}. For instance, consider a translation invariant MPO with $\chi=3$ given by
\begin{equation}
A_m \equiv \left( \begin{array}{ccc}
\mathds{1} & E_m & 0 \\
0 & \lambda \mathds{1} & F_m \\
0 & 0 & \mathds{1}
\end{array} \right),
\label{eq:exampleMPO}
\end{equation}
where $E_m,F_m$ are operators on lattice site $m$ (in our case they will be creation-annihilation operators) and $\lambda$ is a real number. This is the fundamental MPO on which we base all of our constructions. The matrix product of $N$ such matrices will read
\begin{align}
    &A_1 A_2\cdots A_N =\nonumber\\&\left( \begin{array}{ccc}
\mathds{1} & \displaystyle\sum_{1\leq m\leq N} \lambda^{N-m} E_{m} & \displaystyle\sum_{1\leq m<n\leq N} \lambda^{n-m-1} E_m F_n\\[20pt]
0 & \lambda^{N} \mathds{1} & \displaystyle\sum_{1\leq m \leq N} \lambda^{m-1} F_m \\
0 & 0 & \mathds{1}
\end{array} \right).
\end{align}
Choosing $\ket{v}=\ket{1}, \ket{w}=\ket{3}$, we have found an MPO representation for the operator
\begin{equation}
\mc O_{\text{MPO}} = \sum_{1\leq m<n\leq N} \lambda^{n-m-1} E_m F_n.
\label{eq:13corner}
\end{equation}
There is a very elegant way to understand the relation between the operator $\mc O_{\text{MPO}}$ and its MPO representation $(A_m,\ket{v},\ket{w})$ in terms of a \textit{finite state automaton} \cite{CrosswhiteBacon}, i.e. a direct graph where the nodes $i$ correspond to the basis elements $\ket{i}$ of the virtual space and the edges $i\to j$ are labeled by the matrix elements $[A_m]_{ij}$. In our example, the graph looks as follows
\[
\includegraphics[width=0.6\columnwidth]{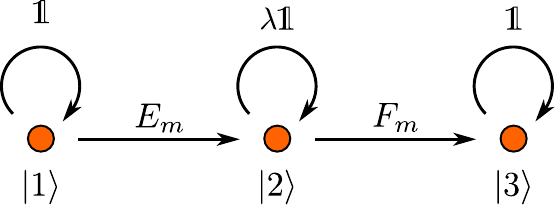}.
\]
We can then assign an operator on the physical space to each path of length $N$ on this graph between the boundary vectors $\ket{v}$ and $\ket{w}$, namely the product of the operators corresponding to each edge on the path, for $m=1,\ldots,N$. The sum of all these terms is the operator $\mc O_{\text{MPO}}$.  All the ``precursor'' MPOs whose continuum limits we care about in this appendix have been obtained from finite state automata in this way. We will not elaborate on this aspect of the construction, but the interested reader can find the corresponding graphs in Fig. \ref{fig:automata}.

A cMPO can be defined by as a limit of discrete MPOs whose matrices $A_i$ have the structure
\begin{equation}
    A_m\equiv \mc I+\epsilon \mc A(x_m)+O(\epsilon^2),
\label{eq:MPOstruct}
\end{equation}
where $\mc I$ is the matrix with only identity operators on the diagonal
\begin{equation}
    \mc I = \left( \begin{array}{ccc}
    \mathds{1} & 0 & \ldots\\
    0 & \mathds{1} & \\
    \vdots & & \ddots
    \end{array} \right),
\end{equation}
and $x_m=m\epsilon$ is a discretization of the real line with lattice parameter $\epsilon$. Then, taking the limits ${N\to\infty}, {\epsilon\to 0}$ while keeping  ${N\epsilon = L}$, we can define the path-ordered exponential
\begin{equation}
\mathcal{P}\exp \left(\int_0^L dx~\mathcal{A}(x) \right) \equiv \lim_{\small{\begin{array}{c} \epsilon\rightarrow 0\\N \rightarrow \infty \end{array}}} \prod_{m=1}^N{\left(\mc I + \epsilon \mc A(x_m)\right)},
\label{defpathexp}
\end{equation}
and the cMPO as one of its matrix elements:
\begin{equation}
\mc O_{\text{cMPO}} \equiv \bra{v}\mathcal{P}\exp \left(\int_0^L dx~\mathcal{A}(x) \right)\ket{w}.
\end{equation}
By taking a doubly infinite series of matrices, and letting $L\to\infty$, cMPOs can be similarly defined on the whole real line.
Now consider the MPO from \eqref{eq:exampleMPO} with
\begin{align}
    E_m &= \dfrac{-i\Lambda}{4}\epsilon\,e^{-\Lambda\epsilon/2} \psi(x_m)\label{def_E}\\
    F_m &= \epsilon\,e^{-\Lambda\epsilon/2} \psi(x_m)\label{def_F}\\
    \lambda&=e^{-\Lambda\epsilon} \label{def_l}
\end{align}
where $\psi(x)$ is the annihilation operator from \eqref{eq:annimagic}. This MPO has the structure \eqref{eq:MPOstruct} for 
\begin{equation}
\mathcal{A}(x) = \left( \begin{array}{ccc}
0 &  \dfrac{-i\Lambda}{4}\psi(x) & 0 \\
0 & -\Lambda \mathds{1}  & \psi(x) \\
0 & 0 & 0
\end{array} \right),
\label{micMPO}
\end{equation}
Using \eqref{eq:13corner} we can find the $(1,3)$ matrix element of the matrix product in \eqref{defpathexp} before taking the limits, which yields
\begin{equation}
    \dfrac{-i\Lambda}{4}\sum_{1\leq m<n\leq N} \epsilon^2 e^{-\Lambda(n-m)\epsilon} \psi(x_m)\psi(x_n).
\end{equation}
Now we take the limits and turn the sums into integrals (notice there is a factor of $\epsilon$ for each variable we sum over). We then find that the cMPO corresponding to \eqref{micMPO} with $\ket{v}=\ket{1},\ket{w}=\ket{3}$ is
\begin{align}
  \mc O_{\text{cMPO}} &=\dfrac{-i\Lambda}{4}\int_{y<x}{dx\,dy\;e^{-\Lambda(x-y)}\psi(x)\psi(y)},\\
  &= \dfrac{-i\Lambda}{8}\int{dx\,dy\;e^{-\Lambda|x-y|}\psi(x)\psi(y)},
\end{align}
which is half of the magic entangler $K$. The other half, the $\psi^\dagger\psi^\dagger$ term, can be obtained analogously, or by taking the Hermitian conjugate. Lastly, the sum of two cMPOs can always be written as a cMPO with bond dimension at most the sum of the two original bond dimensions. In this case, we can write $K$ as a cMPO with $\chi=4$ instead of 6 by defining
\begin{equation}
\mathcal{A}_K(x) \equiv \left( \begin{array}{cccc}
0 &  \dfrac{-i\Lambda}{4}\psi(x) & \dfrac{i\Lambda}{4}\psi^\dagger(x) & 0 \\
0 & -\Lambda \mathds{1}  & 0& \psi(x) \\
0 & 0 & -\Lambda \mathds{1}  & \psi^\dagger(x) \\
0 & 0 & 0 & 0
\end{array} \right).
\label{eq:CFTcMPO}
\end{equation}
and boundary vectors $\ket{v}=\ket{1},\ket{w}=\ket{4}$. It can then be checked that indeed
\begin{equation}
    K = \bra{1}\mathcal{P}\exp \left(\int_{-\infty}^\infty dx~\mathcal{A}_K(x) \right)\ket{4}.
\end{equation}
Thus we are done finding a cMPO representation for the magic entangler $K$. Note that this relied heavily on the fact that the profile of the entangler is exponentially decaying, as opposed to, for instance, a Gaussian (which is the case for the non-magic cMERA presented in the main text).

\subsection{Magic BcMERA}
Let us now check what happens when we make the choice \eqref{eq:magic_g} for the BcMERA from the main text. We can follow the argument in the previous section to find a local parent Hamiltonian for every BcMERA state $\ket{\Psi_B^\Lambda(k,s)}$. Consider the annihilation operators $a_B^\Lambda(k,s)$ that characterize this state 
\begin{equation}
    a_B^\Lambda(k,s)\ket{\Psi_B^\Lambda(k,s)}=0
\end{equation}
These can be written as \eqref{eq:cMERA_anni_expl}, with the momentum fields given by \eqref{Dirichletmode1}-\eqref{Dirichletmode2} (or their Neumann counterparts). We then define the analogous Hamiltonian,
\begin{equation}
    H^\Lambda_B(s)\equiv\int_{0}^\infty{dk\,\varepsilon(k,s)\,{a_B^\Lambda}^\dagger(k,s)a_B^\Lambda(k,s)},
\end{equation}
with $\varepsilon(k,s)$ given once more by \eqref{eq:magic_disp_rel}. Clearly $\ket{\Psi_B^\Lambda(k,s)}$ is the ground state of $H^\Lambda_B(s)$, which in position space reads:
\begin{align}
    H^\Lambda_B(s)&= \ha\int_0^\infty{dx\;\left[\pi(x)^2+(\partial\phi(x))^2\right]}\nonumber\\
    &+\dfrac{1}{2\Lambda^2}\int_0^\infty{dx\;(\partial_x\pi(x))^2}\nonumber\\&+\dfrac{1}{2}\int_0^\infty{dx\;m(s)^2\phi(x)^2},
\end{align}
which is nothing but the restriction of $H^\Lambda(s)$ to the half-line $\R^+$.

We now proceed to verify the cMPS/cMPO compatibility. The dual argument for the UV structure of $\ket{\Psi^\Lambda(s)}$ can be directly imported for $\ket{\Psi_B^\Lambda(s)}$, since we have found the local parent Hamiltonian to be identical to its full-line counterpart. Using the formalism exposed in Appendix \ref{app:cMERA} it is also possible to prove that Eq. \eqref{eq:ndek} holds in the boundary case, restricted to positive momenta.

The cMPO structure of the entangler will take a bit more analysis to pinpoint. Our proposal for the magic BcMERA entangler was made of two pieces. The first one, $K|_{\R^+}$, can be made into a cMPO as in the previous section. The second one, in the language of $\psi(x), \psi^\dagger(x)$, reads
    \begin{equation}
        K_{\text{bdy}}=\dfrac{-i\xi\Lambda}{8}\int{dx\,dy\;e^{\Lambda(x+y)}\psi(x)\psi(y)}+\text{h.c.}
    \end{equation}
It can also be represented as a cMPO, with a rather similar matrix:
\begin{equation}
\mathcal{A}_{\text{bdy}}(x) = \left( \begin{array}{ccc}
-2\Lambda \mathds{1} &  \dfrac{-i\xi\Lambda}{4}\psi(x) & 0 \\
0 & -\Lambda \mathds{1}  & \psi(x) \\
0 & 0 & 0
\end{array} \right),
\end{equation}
and boundary vectors $\ket{v}=\ket{1},\ket{w}=\ket{3}$. Here $\xi=\pm 1$ is defined as in the main text. This yields the $\psi\psi$ term, with an equivalent one for the $\psi^\dagger\psi^\dagger$ one. In fact, we can compress all four terms of the entangler and write it exactly as a cMPO with bond dimension $\chi=5$,
\begin{equation}
    K_{B,R} = \left(\bra{1}+\xi\bra{5}\right)\mathcal{P}\exp \left(\int_{0}^\infty dx~\mathcal{A}_R(x) \right)\ket{4}
\end{equation}
with
\begin{equation}
\mathcal{A}_R(x) \equiv \left(\begin{array}{ccccc}
0 & \dfrac{-i\Lambda}{4}\psi(x) & \dfrac{i\Lambda}{4}\psi^\dagger(x) & 0 & 0 \\
0 & -\Lambda x & 0 & \psi(x) & 0 \\
0 & 0 & -\Lambda x & \psi^\dagger(x) & 0 \\
0 & 0& 0& 0& 0\\
0 & \dfrac{-i\Lambda}{4}\psi(x) & \dfrac{i\Lambda}{4}\psi^\dagger(x) & 0 & -2\Lambda \mathds{1}
\end{array} \right).
\label{bdyAR}
\end{equation}
An equivalent cMPO representation can be found for a BcMERA on the left half-line $\R^-$,
\begin{equation}
    K_{B,L} = \bra{1}\mathcal{P}\exp \left(\int_{-\infty}^0 dx~\mathcal{A}_L(x) \right)\left(\ket{4}+\xi\ket{5}\right)
\end{equation}
with
\begin{equation}
    A_L(x)\equiv\left(\begin{array}{ccccc}
        0 & \dfrac{-i\Lambda}{4}\psi(x) &  \dfrac{i\Lambda}{4}\psi^\dagger(x) & 0 & 0 \\
        0 & -\Lambda\mathds{1} & 0 & \psi(x) & \psi(x) \\
        0 & 0 & -\Lambda\mathds{1} & \psi^\dagger(x) & \psi^\dagger(x) \\
        0 & 0 & 0 & 0 & 0 \\
        0 & 0 & 0 & 0 & -2\Lambda\mathds{1} 
    \end{array}\right).
    \label{bdyAL}
\end{equation}
Note that the boundary vectors corresponding to the physical boundary (resp. $\ket{v}$ and $\ket{w}$) contain a superposition of basis elements that carries the information $\xi$ of the physical boundary conditions.

\subsection{Magic DcMERA}
We proceed analogously to the two previous cases. We denote the annihilation operators that characterize $\ket{\Psi^\Lambda_D(s)}$ by $a_D^\Lambda(k,s)$, and recall that they are given by \eqref{eq:cMERA_anni_expl} with the momentum fields given by \eqref{eq:defectmomfields}. Now we can define
\begin{equation}
    H^\Lambda_D(s)\equiv\int_{0}^\infty{dk\,\varepsilon(k,s)\,{a_B^\Lambda}^\dagger(k,s)a_D^\Lambda(k,s)},
\end{equation}
with $\varepsilon(k,s)$ once again given by \eqref{eq:magic_disp_rel}. By construction the DcMERA state $\ket{\Psi^\Lambda_D(s)}$ is the ground state of $H^\Lambda_D(s)$, which in position space reads
\begin{align}
        H_D^\Lambda(s) &= \ha\int{dx\;\left[\pi(x)^2+(\partial\phi(x))^2\right]}\nonumber\\ &+\dfrac{1}{2\Lambda^2}\int{dx\;(\partial_x\pi(x))^2}\nonumber\\&+ \dfrac{1}{2}\int{dx\;m(s)^2\phi(x)^2},
\end{align}
that is, exactly as $H^\Lambda(s)$, and in particular, it is local. Of course, the two Hamiltonians are not the same, because the fields $\phi(x)$ in each of them belong to different solution spaces of the Klein-Gordon equation. Consequently, they have different ground states.

The fact that $\ket{\Psi^\Lambda_D(s)}$ has a cMPS-compatible UV structure follows exactly as in the previous cases, since its parent Hamiltonian and number density $n(k)$ have exactly the same expressions as in the case without the defect. 

Finally, we proceed to find a cMPO representation for $K_D$. Notice that it can not be given by a translation invariant cMPO, since the presence of the defect breaks translation invariance. It can be seen, however, that we can write it with two $\mc A$ matrices (each at one side of the defect) and an operator insertion $D$ at the position of the defect, keeping bond dimension equal to 5. This should be understood as the continuum limit of a tensor network of this kind:
\[
\includegraphics[width=0.7\columnwidth]{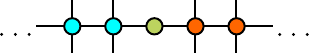}.
\]
Indeed, define
\begin{equation}
    D \equiv \left(\begin{array}{ccccc}
        1 & 0 & 0 &  & -\cos{2\theta} \\
        0 & \sin{2\theta} & 0 & 0 & 0 \\
        0 & 0 & \sin{2\theta} & 0 & 0 \\
        0 & 0 & 0 & 1 & 0 \\
        0 & 0 & 0 & \cos{2\theta} & 0
    \end{array}\right).
\end{equation}
Note $D$ is a matrix on the virtual space of the cMPO whose entries are numbers, instead of physical operators. It thus corresponds to the central tensor, in the diagram, which has no physical legs. It can then be seen that we can express $K_D$ as a cMPO as follows:
\begin{widetext}
\begin{equation}
    K_D = \bra{1}\left[\mathcal{P}\exp\left(\int_{-\infty}^0 dx\;\mathcal{A}_L(x) \right)\right]D\left[\mathcal{P}\exp \left(\int_{0}^\infty dx\;\mathcal{A}_R(x) \right)\right]\ket{4}.
\end{equation}
\end{widetext}
where $\mc A_R$ and $\mc A_L$ are the ones defined in Eqs. \eqref{bdyAR}-\eqref{bdyAL}. 
For totally transmitting defects, $\cos{2\theta}=0$, and we can see that $D$ decouples $\ket{5}$ from the rest of the virtual subspace, so that we can effectively restrict it to $\text{span}\{\ket{1},\ldots\ket{4}\}$, where $\mc A_L(x)= \mc A_R(x)= \mc A(x)$. As expected, for $\theta=\pi/4$, $D$ is trivial and we are back to the usual cMERA, while for $\theta=-\pi/4$, $D$ accounts for the sign changes when crossing the defect. On the other hand, for totally reflecting defects, $\sin{2\theta}=0$, and $D$ factors as
\begin{equation}
    D = \left(\ket{4}\phantom{\big|}+\phantom{\bigg|}\xi_L\ket{5}\right)\bra{4}+\ket{1}\left(\bra{1}\phantom{\big|}+\phantom{\bigg|}\xi_R\bra{5}\right)
\end{equation}
where $\xi_L=1,\xi_R=-1$ for Neumann-Dirichlet ($\theta=0$) and $\xi_L=-1,\xi_R=1$ for Dirichlet-Neumann ($\theta=\frac{\pi}{2}$). This are precisely the combinations of boundary vectors for the two half-lines cMPOs that ensure
\begin{equation}
    K_D = \mathds{1}_L\otimes K_{B,R} + K_{B,L} \otimes \mathds{1}_R,
\end{equation} 
as required for a totally reflecting defect.

In summary, we have seen how the BcMERA and DcMERA built from the magic entangling profile \eqref{eq:magic_g} both satisfy two of the most notable properties of the original magic cMERA, namely the existence of a local parent Hamiltonian $H_B^\Lambda(k,s),H_D^\Lambda(k,s)$, for each cMERA state, and of a cMPO representation for the entangling generators $K_B,K_D$. Regarding this last property, we remark that the principle of minimal updates, while valid, is notexplicitly reflected in the individual matrices $\mc A_L(x), \mc A_R(x)$ (which we could call the ``tensors'' in the continuous tensor network), since they are the same independently of the distance to the boundary/defect. It would be possible, if necessary, to obtain an approximate representation of $K$ that satisfies the minimal update property at the level of the ``tensors'' by truncating away, whenever $|x|\gg\Lambda^{-1}$ the components of $\mc A_L(x),\mc A_R(x)$ that differ from the no-defect tensor $\mc A(x)$, i.e., by restricting them to the $\ket{1},\ldots,\ket{4}$ subspace. For instance, in the defect case, truncating outside some interval $[-x_0, x_0]$, for $\Lambda x_0\gg1$, would be equivalent to inserting a projector $P = \ket{1}\bra{1} + \ldots + \ket{4}\bra{4}$ on the virtual space at the points $\pm x_0$, just as we inserted $D$ at the origin, and replacing $\mc A_L(x), \mc A_R(x)$ by $\mc A(x)$ for $|x|>x_0$. This amounts to restricting the modification to the entangler $K_{\text{def}}$ to have support only within $[-x_0,x_0]$. Alternatively, a smoother truncation could be envisioned in which the components of $\mc A_L(x), A_R(x)$ supported on $\text{span}\{\ket 5\}$ go continuously to zero as $x$ gets further away from the defect (giving up, thus, on the translation invariance of $\mc A_L(x), A_R(x)$).


\bibliography{main}

\begin{thebibliography}{10}

\bibitem{Orus:2013kga}
Roman Orus.
\newblock {A Practical Introduction to Tensor Networks: Matrix Product States
  and Projected Entangled Pair States}.
\newblock {\em Annals Phys.}, 349:117--158, 2014.

\bibitem{Bridgeman:2016dhh}
Jacob~C. Bridgeman and Christopher~T. Chubb.
\newblock {Hand-waving and Interpretive Dance: An Introductory Course on Tensor
  Networks}.
\newblock {\em J. Phys. A}, 50(22):223001, 2017.

\bibitem{tensorsnet}
Glen Evenbly.
\newblock \textit{tensors.net}.
\newblock Available at \url{https://www.tensors.net}.

\bibitem{tensornetworkorg}
Miles Stoudenmire et~al.
\newblock \textit{tensornetwork.org}.
\newblock Available at \url{https://tensornetwork.org/contribute/}.

\bibitem{Vidal:2007hda}
G.~Vidal.
\newblock {Entanglement Renormalization}.
\newblock {\em Phys. Rev. Lett.}, 99(22):220405, 2007.

\bibitem{Vidal:2008zz}
G.~Vidal.
\newblock {Class of Quantum Many-Body States That Can Be Efficiently
  Simulated}.
\newblock {\em Phys. Rev. Lett.}, 101:110501, 2008.

\bibitem{Aguado:2007oza}
Miguel Aguado and Guifr\'e Vidal.
\newblock {Entanglement renormalization and topological order}.
\newblock {\em Phys. Rev. Lett.}, 100:070404, 2008.

\bibitem{FerrisPoulin_2014}
Andrew~J. Ferris and David Poulin.
\newblock Tensor networks and quantum error correction.
\newblock {\em Physical Review Letters}, 113(3), Jul 2014.

\bibitem{beny2013deep}
Cédric Bény.
\newblock Deep learning and the renormalization group, 2013.

\bibitem{Swingle_2012}
Brian Swingle.
\newblock Entanglement renormalization and holography.
\newblock {\em Physical Review D}, 86(6), Sep 2012.

\bibitem{Czech:2015kbp}
Bartlomiej Czech, Lampros Lamprou, Samuel McCandlish, and James Sully.
\newblock {Tensor Networks from Kinematic Space}.
\newblock {\em JHEP}, 07:100, 2016.

\bibitem{VerstraeteCirac}
F.~Verstraete and J.~I. Cirac.
\newblock Continuous matrix product states for quantum fields.
\newblock {\em Phys. Rev. Lett.}, 104:190405, May 2010.

\bibitem{Haegeman:2011uy}
Jutho Haegeman, Tobias~J. Osborne, Henri Verschelde, and Frank Verstraete.
\newblock {Entanglement Renormalization for Quantum Fields in Real Space}.
\newblock {\em Phys. Rev. Lett.}, 110(10):100402, 2013.
\newblock \textit{Note}: There is a lot of material in the appendices of
  arXiv:1102.5524v1, which was not included in the second version or the
  published version.

\bibitem{Jennings:2015nwa}
David Jennings, Jutho Haegeman, Tobias~J. Osborne, and Frank Verstraete.
\newblock {Continuum tensor network field states, path integral representations
  and spatial symmetries}.
\newblock {\em New J. Phys.}, 17(6):063039, 2015.

\bibitem{Tilloy:2018gvo}
Antoine Tilloy and J.~Ignacio Cirac.
\newblock {Continuous Tensor Network States for Quantum Fields}.
\newblock {\em Phys. Rev. X}, 9(2):021040, 2019.

\bibitem{Hu:2018hyd}
Qi~Hu, Adrian Franco-Rubio, and Guifr\'e Vidal.
\newblock {Continuous tensor network renormalization for quantum fields}.
\newblock 8 2018.

\bibitem{Franco-Rubio:2017tkt}
Adri\'an Franco-Rubio and Guifr\'e Vidal.
\newblock {Entanglement and correlations in the continuous multi-scale
  entanglement renormalization ansatz}.
\newblock {\em JHEP}, 12:129, 2017.

\bibitem{Franco-Rubio:2019nne}
Adri\'an Franco-Rubio and Guifr\'e Vidal.
\newblock {Entanglement renormalization for gauge invariant quantum fields}.
\newblock {\em Phys. Rev. D}, 103(2):025013, 2021.

\bibitem{Hu:2017rsp}
Qi~Hu and Guifr\'e Vidal.
\newblock {Spacetime Symmetries and Conformal Data in the Continuous Multiscale
  Entanglement Renormalization Ansatz}.
\newblock {\em Phys. Rev. Lett.}, 119(1):010603, 2017.

\bibitem{Nozaki:2012zj}
Masahiro Nozaki, Shinsei Ryu, and Tadashi Takayanagi.
\newblock {Holographic Geometry of Entanglement Renormalization in Quantum
  Field Theories}.
\newblock {\em JHEP}, 10:193, 2012.

\bibitem{Mollabashi:2013lya}
Ali Mollabashi, Masahiro Nozaki, Shinsei Ryu, and Tadashi Takayanagi.
\newblock {Holographic Geometry of cMERA for Quantum Quenches and Finite
  Temperature}.
\newblock {\em JHEP}, 03:098, 2014.

\bibitem{Miyaji:2014mca}
Masamichi Miyaji, Shinsei Ryu, Tadashi Takayanagi, and Xueda Wen.
\newblock {Boundary States as Holographic Duals of Trivial Spacetimes}.
\newblock {\em JHEP}, 05:152, 2015.

\bibitem{Molina-Vilaplana:2015mja}
Javier Molina-Vilaplana.
\newblock {Information Geometry of Entanglement Renormalization for free
  Quantum Fields}.
\newblock {\em JHEP}, 09:002, 2015.

\bibitem{Chapman_2018}
Shira Chapman, Michal~P. Heller, Hugo Marrochio, and Fernando Pastawski.
\newblock Toward a definition of complexity for quantum field theory states.
\newblock {\em Physical Review Letters}, 120(12), Mar 2018.

\bibitem{Evenbly2010}
G.~Evenbly, R.~N.~C. Pfeifer, V.~Picó, S.~Iblisdir, L.~Tagliacozzo, I.~P.
  McCulloch, and G.~Vidal.
\newblock Boundary quantum critical phenomena with entanglement
  renormalization.
\newblock {\em Physical Review B}, 82(16), Oct 2010.

\bibitem{Evenbly2014}
G.~Evenbly and G.~Vidal.
\newblock Algorithms for entanglement renormalization: Boundaries, impurities
  and interfaces.
\newblock {\em Journal of Statistical Physics}, 157(4-5):931–978, Apr 2014.

\bibitem{Evenbly:2013tta}
Glen Evenbly and Guifr\'e Vidal.
\newblock {A theory of minimal updates in holography}.
\newblock {\em Phys. Rev. B}, 91:205119, 2015.

\bibitem{Hauru:2015abi}
Markus Hauru, Glen Evenbly, Wen~Wei Ho, Davide Gaiotto, and Guifr\'e Vidal.
\newblock {Topological conformal defects with tensor networks}.
\newblock {\em Phys. Rev. B}, 94(11):115125, 2016.

\bibitem{Bridgeman:2017etx}
Jacob~C. Bridgeman and Dominic~J. Williamson.
\newblock {Anomalies and entanglement renormalization}.
\newblock {\em Phys. Rev. B}, 96(12):125104, 2017.

\bibitem{Czech:2016nxc}
Bartlomiej Czech, Phuc~H. Nguyen, and Sivaramakrishnan Swaminathan.
\newblock {A defect in holographic interpretations of tensor networks}.
\newblock {\em JHEP}, 03:090, 2017.

\bibitem{Chapman:2018bqj}
Shira Chapman, Dongsheng Ge, and Giuseppe Policastro.
\newblock {Holographic Complexity for Defects Distinguishes Action from
  Volume}.
\newblock {\em JHEP}, 05:049, 2019.

\bibitem{Zou:2019xbi}
Yijian Zou, Martin Ganahl, and Guifr\'e Vidal.
\newblock {Magic entanglement renormalization for quantum fields}.
\newblock 6 2019.

\bibitem{Bachas:2001vj}
C.~Bachas, J.~de~Boer, R.~Dijkgraaf, and H.~Ooguri.
\newblock {Permeable conformal walls and holography}.
\newblock {\em JHEP}, 06:027, 2002.

\bibitem{Hung:2021tsu}
Ling-Yan Hung and Guifr\'e Vidal.
\newblock {Continuous entanglement renormalization on the circle}.
\newblock 1 2021.

\bibitem{RevModPhys.47.773}
Kenneth~G. Wilson.
\newblock The renormalization group: Critical phenomena and the kondo problem.
\newblock {\em Rev. Mod. Phys.}, 47:773--840, Oct 1975.

\bibitem{evenbly2013quantum}
Glen Evenbly and Guifr\'e Vidal.
\newblock Quantum criticality with the multi-scale entanglement renormalization
  ansatz, 2013.

\bibitem{Cotler:2018ehb}
Jordan~S. Cotler, M.~Reza Mohammadi~Mozaffar, Ali Mollabashi, and Ali Naseh.
\newblock {Entanglement renormalization for weakly interacting fields}.
\newblock {\em Phys. Rev. D}, 99(8):085005, 2019.

\bibitem{Cotler:2018ufx}
Jordan Cotler, M.~Reza Mohammadi~Mozaffar, Ali Mollabashi, and Ali Naseh.
\newblock {Renormalization Group Circuits for Weakly Interacting Continuum
  Field Theories}.
\newblock {\em Fortsch. Phys.}, 67(10):1900038, 2019.

\bibitem{Fernandez-Melgarejo:2019sjo}
Jose~J. Fernandez-Melgarejo, Javier Molina-Vilaplana, and Emilio
  Torrente-Lujan.
\newblock {Entanglement Renormalization for Interacting Field Theories}.
\newblock {\em Phys. Rev. D}, 100(6):065025, 2019.

\bibitem{Fernandez-Melgarejo:2020fzw}
Jose~J. Fernandez-Melgarejo and Javier Molina-Vilaplana.
\newblock {Non-Gaussian Entanglement Renormalization for Quantum Fields}.
\newblock {\em JHEP}, 07:149, 2020.

\bibitem{numintercMERA}
Yijian Zou, Martin Ganahl, and Guifr\'e Vidal.
\newblock {Numerical results for interacting magic cMERA}.
\newblock In preparation.

\bibitem{ReedSimon}
Michael Reed and Barry Simon.
\newblock {\em Methods of Modern Mathematical Physics. Vol. II: Fourier
  Analysis, Self-Adjointness}.
\newblock Academic Press, New York, 1975.

\bibitem{Asorey:2004kk}
M.~Asorey, A.~Ibort, and G.~Marmo.
\newblock {Global theory of quantum boundary conditions and topology change}.
\newblock {\em Int. J. Mod. Phys. A}, 20:1001--1026, 2005.

\bibitem{Asorey:2015lja}
M.~Asorey, D.~Garc\'\i{}a-Alvarez, and J.~M. Mu\~noz Casta\~neda.
\newblock {Boundary Effects in Bosonic and Fermionic Field Theories}.
\newblock {\em Int. J. Geom. Meth. Mod. Phys.}, 12(06):1560004, 2015.

\bibitem{Cardy:1984bb}
John~L. Cardy.
\newblock {Conformal Invariance and Surface Critical Behavior}.
\newblock {\em Nucl. Phys. B}, 240:514--532, 1984.

\bibitem{Haegeman:2010fv}
Jutho Haegeman, J.~Ignacio Cirac, Tobias~J. Osborne, Henri Verschelde, and
  Frank Verstraete.
\newblock {Applying the variational principle to (1+1)-dimensional quantum
  field theories}.
\newblock {\em PoS}, FACESQCD:029, 2010.

\bibitem{CrosswhiteBacon}
Gregory~M. Crosswhite and Dave Bacon.
\newblock Finite automata for caching in matrix product algorithms.
\newblock {\em Phys. Rev. A}, 78:012356, Jul 2008.

\end{thebibliography}
\bibliographystyle{unsrt}

\end{document}